\documentclass{aa}  
\usepackage{longtable}
\usepackage{caption}
\usepackage{subcaption}
\usepackage{amsmath}
\usepackage{graphicx}
\usepackage{float}
\usepackage{txfonts}
\usepackage[fleqn]{amsmath}	
\usepackage{amssymb}	
\usepackage{mathtools}
\usepackage{hyperref}
\hypersetup{
  colorlinks=true,   
  citecolor=blue,    
  urlcolor=blue,     
  linkcolor=blue,     
}
\bibpunct{(}{)}{;}{a}{}{,} 
\usepackage{lscape}

\newcommand{\oiii}{[\ion{O}{iii}]\xspace}
\newcommand{\Hb}{H$\beta$\xspace}
\newcommand{\Ha}{H$\alpha$\xspace}
\newcommand{\Mgii}{\ion{Mg}{ii}\xspace}

\newcommand{\Feii}{\ion{Fe}{ii}\xspace}

\newcommand{\Mbh}{$M_{\rm BH}$\xspace}
\newcommand{\Lad}{$L_{\rm AD}$\xspace}
\newcommand{\ER}{$\lambda_{\rm Edd}$\xspace}

\newcommand{\kbol}{$k_{\rm bol}$\xspace}

\DeclareRobustCommand{\ion}[2]{%
\relax\ifmmode
\ifx\testbx\f@series
{\mathbf{#1\,\mathsc{#2}}}\else
{\mathrm{#1\,\mathsc{#2}}}\fi
\else\textup{#1\,{\mdseries\textsc{#2}}}%
\fi}

\newcommand{\jwst}{\textit{JWST}\xspace}

\begin{document}

   \title{The accretion of quasars at the epoch of reionisation: \textit{JWST} catches the primeval monsters slowly feasting}

  \author{Bartolomeo Trefoloni\inst{1,2}
  \thanks{\email{bartolomeo.trefoloni@sns.it}},
  Emanuele~Nardini\inst{2},
  Stefano~Carniani\inst{1},
  Elisabeta~Lusso\inst{2,3},
  Alessandro~Marconi\inst{2,3},
  Eleonora~Parlanti\inst{1},
  Andrea~Sacchi\inst{4,5},
  Anastasia~Shlentsova\inst{2,3,6},
  Matilde~Signorini\inst{3,7},
  Guido~Risaliti\inst{2,3},
  Sandra~Zamora\inst{1}
  }

\institute{
$^{1}$Scuola Normale Superiore, Piazza dei Cavalieri 7, I-56126 Pisa, Italy\\
$^{2}$INAF -- Osservatorio Astrofisico di Arcetri, Largo Enrico Fermi 5, I-50125 Firenze, Italy\\
$^{3}$Dipartimento di Fisica e Astronomia, Universit\`a degli Studi di Firenze, via G. Sansone 1, 50019 Sesto Fiorentino, Firenze, Italy\\
$^{4}$INAF - Istituto di Astrofisica Spaziale e Fisica Cosmica Milano, Via A.Corti 12, 20133 Milano, Italy\\
$^{5}$Center for Astrophysics $\vert$ Harvard \& Smithsonian, 60 Garden Street, Cambridge, MA 02138, USA\\
$^{6}$Instituto de Astrof\'isica, Facultad de F\'isica, Pontificia Universidad Cat\'olica de Chile, Casilla 306, Santiago 22, Chile\\
$^{7}$European Space Agency (ESA), European Space Research and Technology Centre (ESTEC), Keplerlaan 1, 2201 AZ Noordwijk, The Netherlands}

\titlerunning{The accretion of quasars at the epoch of reionisation}
\authorrunning{B. Trefoloni et al.}

\abstract{Quasars (QSOs) emit an enormous amount of light as a result of the accretion of gas onto supermassive black holes (SMBHs). Thanks to their luminosity, the most distant known QSOs allow us to trace the growth of SMBHs deep into the epoch of reionisation. In this work, we employed \textit{JWST}/NIRSpec observations of eight luminous ($\log(L_{3000\,\AA}/(\rm erg \, s^{-1}))$\,$>$\,45.7) QSOs at $z$\,$\geq$\,5.9 to constrain their accretion properties, namely black hole mass, accretion disc (AD) luminosity, and Eddington ratio ($M_{\rm BH}$, $L_{\rm AD}$, $\lambda_{\rm Edd}$), by fitting the rest-frame UV and optical emission with different AD models. This method provided self-consistent measurements of both $M_{\rm BH}$ and $L_{\rm AD}$. The uncertainties on $M_{\rm BH}$ and $L_{\rm AD}$, obtained within the AD-modelling framework ($\sigma^{\rm AD}_{M_{\rm BH}}$\,$\sim$\,0.2 dex;  $\sigma^{\rm AD}_{L_{\rm AD}}$\,$\sim$\,0.1 dex), are significantly smaller than the systematic uncertainties associated with single-epoch $M_{\rm BH}$ ($\sim$0.4 dex) and $L_{\rm AD}$ derived via bolometric corrections ($\sim$0.2 dex). Based on these results, in our sample we found an average Eddington ratio of $\langle \log(\lambda_{\rm Edd}) \rangle$\,=\,$-0.9$, with a dispersion of $\sim$\,0.2 dex. Assuming that our high-$z$ QSOs are representative of optically-selected bright blue QSOs, we derive a fraction of systems accreting above the Eddington limit of $\sim$\,0.2\%. In conclusion, this work i) demonstrates the suitability of \textit{JWST} to test AD models on high-redshift ($z$\,$\gtrsim$\,4) QSOs, thanks to the large NIRSpec spectral coverage; ii) shows that AD modelling can yield robust $M_{\rm BH}$ and $L_{\rm AD}$ measurements, with smaller uncertainties than the typical calibrations; and iii) provides compelling evidence for sub-Eddington accretion in bright high-$z$ QSOs, challenging the widespread paradigm of near- or super-Eddington accretion occurring in these sources.}


   \maketitle
%

\section{Introduction}

Accretion of gas onto a supermassive black hole converts gravitational energy into radiation. If this process, occurring through an accretion disc (AD), is efficient, the huge energetic output released in the nuclear region of a galaxy gives rise to an active galactic nucleus (AGN), with quasars (QSOs) belonging to the brightest end of the AGN luminosity distribution.

Because of their incidence and brightness, broad-line AGN can be observed up to extremely high redshifts ($z$\,$\geq$\,8.5; e.g., \citealt{kokorev2023uncover, taylor2025capers, juodvzbalis2025jades}), possibly as high as z$\sim$11 where the direct detection of broad lines in GN-z11 (\citealt{maiolino2024small}) is still debated. At the same time, the farthest QSOs have been detected at $z$\,$\sim$\,7.5--7.6 (\citealt{banados2018, wang2021luminous}), when the Universe was only 0.7 Gyr old. The presence of SMBHs with masses exceeding $10^{8} \, M_{\odot}$ at so early cosmic times remains puzzling, as the pathway to grow such high masses in so little time is largely unclear (see, e.g., \citealt{inayoshi2020assembly, lusso2022dawn, fan2023quasars}, for recent reviews on this topic). In particular, it raises compelling questions concerning both the mechanisms at work for this early assembly of SMBHs and the reliability of their mass measurements.

Reverberation mapping (RM; \citealt{blandford1982reverberation}) is widely recognized as the benchmark technique to estimate BH masses in AGN, although other approaches have also been explored in recent years (see, e.g., \citealt{gliozzi2024comparing} for a comparison). The RM method relies on the delay between the optical AD continuum emission and the response of broad emission lines (generally \Hb; e.g., \citealt{peterson04}). Assuming that the clouds orbiting in the broad-line region (BLR) are virialised, it is ultimately possible to infer the BH mass based on the BLR radius and cloud velocity. 
However, RM is extremely time- and resource-consuming, and can only be applied to broad-line AGN with significant continuum variations on relatively short time-scales (see, e.g., \citealt{shen2019sloan}, and references therein). Because of these limitations, the number of AGN where RM has been successfully performed is of the order of a few hundreds (e.g., \citealt{shen2024sloan}), yet many more are expected in the next years as part of the on-going `Black hole mapper' programme in the recently started Sloan Digital Sky Survey V (SDSS-V; \citealt{kollmeier2019sdss, almeida2023eighteenth}).

A more `economic' BH-mass measurement method is based on the tight relation, with a scatter of $<$\,0.2 dex (\citealt{kaspi2005relationship, bentz2009radius}), between the BLR radius and the monochromatic disc luminosity (the $R$--$L$ relation; \citealt{kaspi2000reverberation}). Such a discovery allowed the use of the monochromatic luminosity -- or, equivalently, the line luminosity -- as a proxy for the BLR radius, enabling the calibration of `single-epoch' (SE) black-hole mass measurements, whereby a single spectrum can, in practice, provide an estimate of both the key ingredients to derive \Mbh, namely the BLR radius and the rotational velocity (parameterized by the second moment of the line profile or the full width at half maximum -- FWHM; e.g., \citealt{vestergaard2006determining}, and references therein). Consequently, SE calibrations have been widely employed to estimate the BH masses of moderately luminous broad-line AGN, as well as of bright QSOs (e.g., \citealt{shen2011, rakshit2020spectral, wu2022catalog}). Despite their wide applicability and convenience, non-negligible systematic uncertainties affect the resulting BH mass estimates (see, e.g., \citealt{shen2013mass}, and Appendix \ref{app:systematics} for a more detailed discussion). Examples of such uncertainties are the unknown geometry of the BLR (key to reliably estimate \Mbh), the inclination of the line of sight to the BLR, the differential line/continuum variability, and the non-Gaussianity of the line profile, just to mention some. The complicated dynamics of the BLR, possibly also affected by a non-virialised component (i.e., an outflow; \citealt{pancoast2014modelling}), and the effect of radiation pressure on the BLR clouds (e.g., \citealt{marconi2008effect}), exacerbated in the case of bright sources (e.g., \citealt{fries2024sdss}), are additional reasons for concern. The combination of all these limitations led to even question the extrapolation of SE calibrations to the luminosity and redshift regimes of bright QSOs (\citealt{bertemes2025jwst}).

While other techniques, such as dynamical methods (e.g. \citealt{ferrarese2005supermassive}) are practically unfeasible for large samples at high redshift (but see also \citealt{juodvzbalis2025direct} for a direct BH mass measurement in an AGN at $z\sim$7), accretion-disc modelling offers an alternative approach to self-consistently estimate the accretion parameters underlying the QSO emission (e.g., \citealt{malkan1983ultraviolet}). Indeed, the disc steady-state spectral energy distribution (SED) is mainly governed by a few parameters, namely the black-hole mass (\Mbh), the accretion rate ($\dot{M}$), and the accretion efficiency ($\eta$), which is related to the spin $a^{*}$ (e.g., \citealt{shakura1973black, novikov1973astrophysics, page1974disk}). The combination of the latter parameters gives the AD luminosity as $L_{\rm AD}$\,=\,$
\eta(a^{*}) \dot{M} c^2$. While fitting the observed SED with AD models entails the obvious advantage of simultaneously constraining both \Mbh and \Lad, this approach has been generally attempted only on relatively small samples with remarkable data quality (e.g., \citealt{capellupo2015active, campitiello2020estimating, wolf2024accretion}). This is mostly due to the demanding requirements needed to reliably describe the disc SED. It is necessary to sample the SED with adequate signal-to-noise (S/N) near its peak, while covering -- ideally simultaneously -- the largest possible wavelength range. The presence of other estimates of \Mbh and \Lad from the SE relations and bolometric corrections can also provide collateral constraints on the models.

The new capabilities offered by \jwst (\citealt{gardner2006james}), revolutionised our knowledge of the early Universe. In particular, the near-IR coverage delivered by the NIRSpec instrument (0.6--5.3 $\mu$m; \citealt{boker2022near, jakobsen2022near}) offered, for the first time, the possibility to observe the \Hb--\oiii complex with unprecedented sensitivity up to $z$\,$\sim$\,9. Recently, \citet{trefoloni2025_ganifs} showed that, leveraging the \jwst low-resolution observing mode with the PRISM/CLEAR coupling, it is possible to employ AD models to robustly constrain the accretion properties of high-redshift QSOs. Following their approach, in this paper we analyse a sample of bright QSOs at $z$\,$\gtrsim$\,5.9 observed across the full \jwst/NIRSpec wavelength range, to investigate their accretion properties and to critically compare the robustness of SE- and AD-based results. 

Here, we briefly outline the structure of the paper. In Section \ref{sec:data}, we describe the dataset employed, while in Section \ref{sec:methods} we explain the analyses performed. We describe our results in Section \ref{sec:results}. We discuss our findings in Section \ref{sec:discussion} in the context of other similar works, and summarise them in Section \ref{sec:conclusions}. Throughout this paper we adopt a flat $\Lambda$CDM cosmology with $H_0 = 70$ km s$^{-1}$ Mpc$^{-1}$, $\Omega_{\Lambda}$ = 0.7, and $\Omega_{\rm m}$ = 0.3.

\section{Data}
\label{sec:data}
\begin{figure}[h!]
\centering
\includegraphics[width=\linewidth,clip]{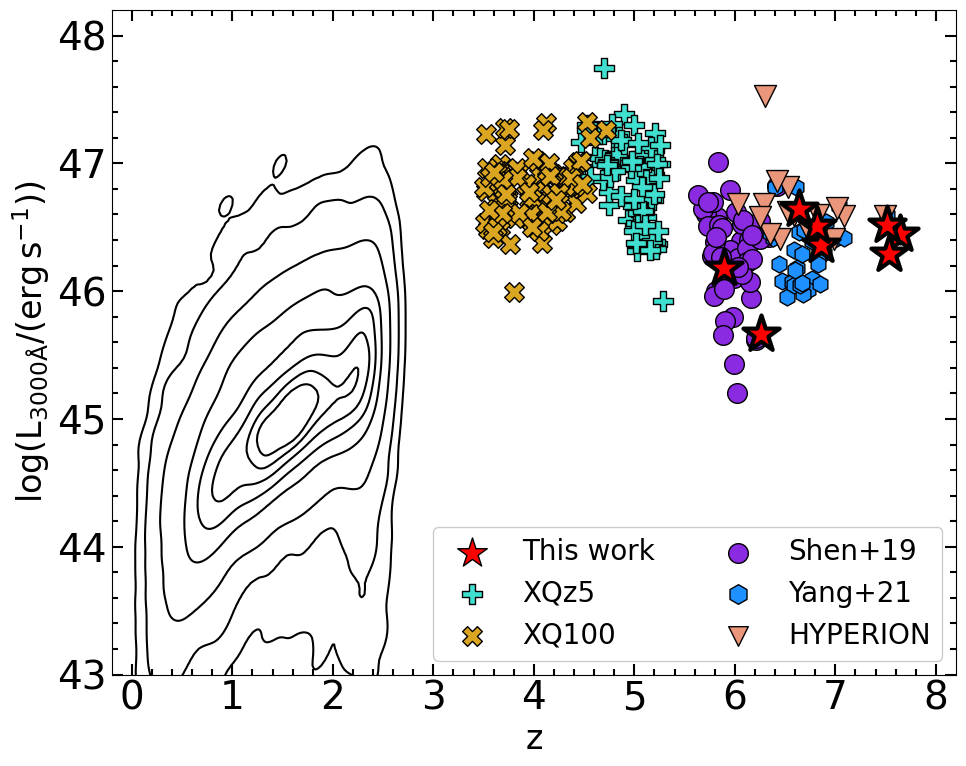}
\caption{QSOs presented in this work (red stars) and other QSO surveys at different redshifts, namely XQz5 \citep{lai2024xqz5}, ZQ100 \citep{lopez2016}, the Gemini/GNIRS sample of \citet{shen2019gemini}, the $z$\,$\sim$\,6.3--7.6 sample of \citet{yang2021probing}, and HYPERION \citep{zappacosta2023hyperluminous}. 
Quasars at $z$\,$\lesssim$\,2.6 from the SDSS DR16Q (\citealt{wu2022catalog}) are shown as black contours. Some high-$z$ QSOs are shared among multiple surveys.}
\label{fig:L3000_z}
\end{figure}

Our two-prong approach to estimate the accretion parameters of high-redshift ($z$\,$\geq$\,5.9) QSOs observed with \jwst requires the widest possible wavelength coverage. Specifically, we focused on datasets that include at least two robustly detected broad emission lines with available SE calibrations (but avoiding \ion{C}{iv}\,$\lambda$1549, as we explain in Sec. \ref{sec:SE_MBH}). At the same time, in order to anchor the AD emission to as many continuum windows as possible, a spectral coverage from the Ly$\alpha$ to the \Ha is desirable. Hence, here we assembled a compilation of 8 QSOs with publicly available spectroscopic data obtained with \jwst/NIRSpec, probing the rest-frame range between $\sim$\,1,200--6,700 \AA. At the redshifts covered by our sample (i.e., well into the reionisation epoch), the spectral region bluewards of the Ly$\alpha$ becomes inaccessible due to the almost total absorption by hydrogen in the intergalactic medium (IGM). General information about the sources, as well as the reference works where their \jwst observations are presented, are provided in Table \ref{tbl:gen_info}. We refer to those works for the details about the observing set-ups and the data reduction. For the sources whose \textit{JWST} observations have not been published yet, we used the data available in the Barbara A. Mikulski Archive for Space Telescopes (MAST), adopting the standard calibration pipeline (\citealt{bushouse2023jwst}). We emphasize that our sample is rather various and unbiased, as a result of the heterogeneous selection criteria that motivated the observations and of the different science cases that these data were meant to address. 
We present the selected QSOs together with other sources observed in several other surveys in Fig. \ref{fig:L3000_z}.

\begin{table*}[h!]
\caption{Relevant information for the objects in our sample.}
\centering
\setlength{\tabcolsep}{5pt}
\begin{tabular}{c c c c c c}
 \hline \noalign{\smallskip}
 Source & RA  & DEC & $z$ & $ \log(L_{3000\,\AA} /(\rm erg \, s^{-1}))$  & Ref.\\
    
 \hline \noalign{\smallskip}
 J0020$-$3653   &  5.131   & $-$36.895  & 6.860 & 46.36 & \citet{christensen2023metal} \\ 
 J0313$-$1806   &  48.433  & $-$18.110  & 7.642 & 46.45 & \citet{wang2021luminous} \\ 
 J0411$-$0907   &  62.869  & $-$9.131   & 6.825 & 46.51 & \citet{christensen2023metal} \\ 
 J0910$-$0414   &  137.727 & $-$4.235   & 6.636 & 46.64 & \citet{wang2019exploring} \\ 
 J1007$+$2115   &  151.993 & 21.258  & 7.515 & 46.51 & \citet{yang2020poniua} \\ 
 J1342$+$0928   &  205.534 & 9.477  & 7.535 & 46.29 & \citet{christensen2023metal} \\ 
 J1425$+$3254   &  216.318 & 32.903 & 5.890 & 46.18 & \citet{marshall2025jwst} \\ 
 J2239$+$0207   &  339.948 & 2.130  & 6.260 & 45.66 & \citet{lyu2025fading} \\ 

 \hline
\end{tabular}
\label{tbl:gen_info}
\tablefoot{The Ref. column reports the works presenting the observations considered here. In case no papers concerning these observations have been produced yet, we report the discovery paper.}
\end{table*}

In the cases where we extracted the nuclear spectrum from the IFU data-cube (namely J0910$-$0414, J1425+3254 and J2239+0207), the radius of the integration region was chosen to be $1\farcs0$, therefore virtually including all the flux enclosed within the PSF, whose typical size is of the order of $0\farcs1$ (see, e.g., \citealt{deugenio2024fast}). In the extraction, with the aim of producing a more realistic estimate of the flux uncertainties, we rescaled the formal uncertainty on the integrated spectrum based on the `ERR' extension using the flux standard deviation in small ($\sim$\,20--30 \AA) continuum windows free from emission lines. This allowed us to take into account the correlations induced by the size of the PSF relative to the spaxel size (see, e.g., \citealt{ubler2023ga}).

\section{Methods}
\label{sec:methods}

\subsection{Spectral fits}
\label{sec:blr_fit}

With the aim of deriving the spectral properties of the most widely used virial emission lines (\Mgii, \Hb, \Ha)\footnote{Calibrations involving the \ion{C}{iv}\,$\lambda$1549 line are still quite uncertain (see, e.g., \citealt{coatman2017} for an overview on this topic and the attempt to correct SE \ion{C}{iv}-based \Mbh values), mostly because of the difficulty in separating the virial from the outflow component. Because of this, we refrained from employing this line.} and the nearby continuum, we performed a spectral fit of the nuclear spectra for all our sources. We achieved this by employing a custom-made Python code, based on the IDL \textsc{MPFIT} package \citep{Markwardt2009}, which takes advantage of the Levenberg-Marquardt technique \citep{more1978levenberg} to solve the least-squares problem.

Broad lines often exhibit complex morphologies, which can hardly be described by simple analytical functions. The overall line profile indeed retains a larger amount of physical information about the BLR kinematics, traced by spectral signatures such as asymmetries, double peaks, and non-Gaussian tails, rather than a single parameter representative of the line width (e.g., \citealt{flohic2012effects, kollatschny2013shape, storchi2017double}). Also, several SE calibrations call for the second moment of the line profile ($\sigma$) to be a more reliable proxy of the BLR velocity (see Sec.~1.2 in \citealt{db2020}) than the mere FWHM. For this reason, in order to model the broad line profiles more faithfully, we adopted a two-stage approach. First, we considered a set of emission line profiles (Gaussians, Lorentzians, or broken power laws with a Gaussian smoothing kernel) to reproduce as best as possible both the narrow (FWHM\,$<$\,1,000 km s$^{-1}$) and the broad lines (FWHM\,$\geq$\,1,000 km s$^{-1}$), also including the broad \ion{Fe}{ii} emission and possible outflow components. We modelled the \ion{Fe}{ii} and the narrow lines (\Ha, \Hb, [\ion{O}{iii}], with the kinematics of the latter two emission lines constrained to be the same) in the same way as described in several other previous works \citep{trefoloni2023most,trefoloni2024missing,trefoloni2025_ganifs}, to which we refer the interested reader for a more thorough description of the set-up. We chose different spectral ranges to fit the line complexes close to the emission lines of interest. The fit was performed roughly in the range 2,200--3,500~\AA\ for the \Mgii line, between 4,200--5,500~\AA\ for the \Hb, and between 6,200--6,900~\AA\ for the \Ha. 

\begin{figure*}[h!]
\centering
\includegraphics[width=\linewidth,clip]{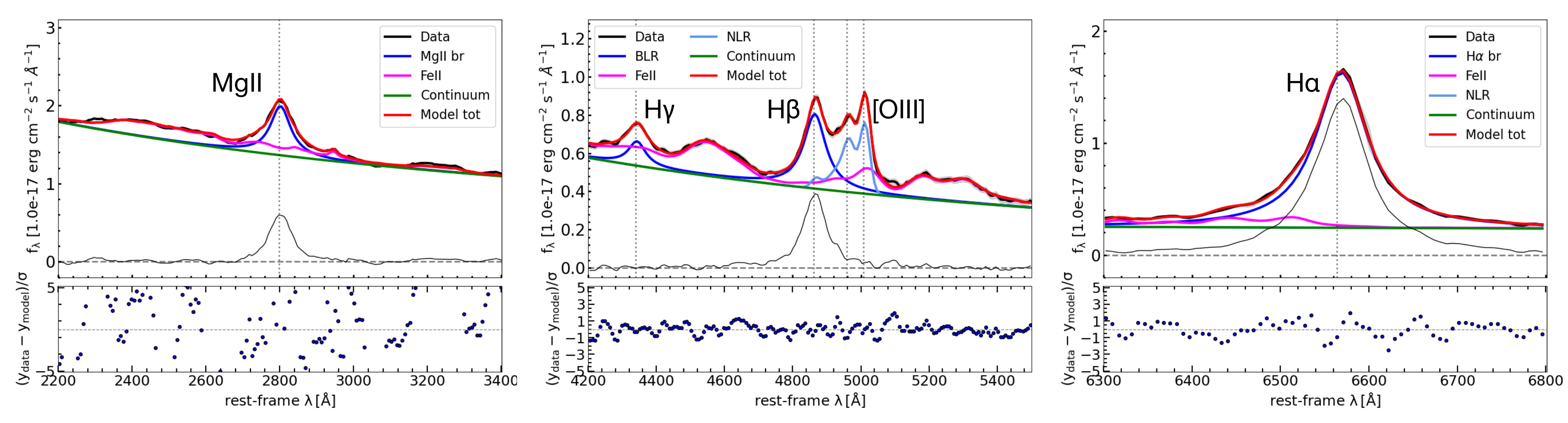}
\caption{Spectral fits of \Mgii, \Hb, and \Ha lines of J1425$+$3254. Fluxes are shown in the rest frame. All the components are colour-coded according to the legend. The remaining spectrum after the best-fit model subtraction is shown as a black line. The dotted lines show the expected wavelengths for the most prominent emission lines. For the \Mgii line here we only show the best fit obtained assuming a Lorentzian profile.}
\label{fig:ex_fits}
\end{figure*}

After the best-fit model was derived, we subtracted all the best-fit components with the exclusion of the broad line of interest (\Mgii, \Hb, or \Ha, see black line in Fig. \ref{fig:ex_fits}). This approach proved successful in isolating the broad-line emission in both \Hb and \Ha. In the case of the \Mgii line, the separation between the broad and the narrow component is often unclear, especially in the low-resolution data. Because of this, we performed the fit of this line adopting both a double Gaussian decomposition (one broad and one narrow component) and a single Lorentzian. When estimating \Mbh using the \Mgii calibrations, we present the results associated with both decompositions. We show an example of the best-fit model as well as the individual components in Fig.~\ref{fig:ex_fits}, while all the others are presented in Appendix \ref{app:fit_atlas}.

At the end of this `semi-parametric' approach, we used the broad-line profile to measure the line parameters. We considered the line emission between $\pm$\,10,000 km s$^{-1}$ from the expected peak location and directly computed the flux, the first and second ($\sigma$) line moments, and the FWHM. Here, we also performed a spectral smoothing adopting a Gaussian kernel 400 $\rm km \, s^{-1}$ wide, in order to avoid noise spikes when computing the line parameters. The width of this kernel is much smaller than the line width, and therefore does not affect the line $\sigma$, nor the FWHM. Performing the same analysis, using the wavelengths corresponding to the $1^{\rm st}$ and the $99^{\rm th}$ percentiles of the line flux distribution instead, would produce negligible differences in the line parameters.

We estimated the uncertainties on the parameters of interest by producing 100 mock model-subtracted spectra by randomising the flux in each spectral channel. Here we neglected the uncertainty introduced by the subtraction of the best-fit model components. However, we stress that the ultimate goal of this procedure is to provide us with estimates of the line parameters (luminosity, width) or continuum luminosities. Since the typical uncertainties on these quantities are much smaller than the systematic uncertainties on the calibrations where we employ them ($\lesssim$\,0.1 dex against $\sim$\,0.4 dex), we can safely neglect this additional source of uncertainty. We report the relevant parameters derived from the fit, together with the relative uncertainties, in Table \ref{tbl:spec_pars}. We note that, for the \Mgii line of J0910$-$0414, we took advantage of the line measurement already performed by \citet{yang2021probing}. There, the authors fitted a Gaussian profile to the data obtained with GNIRS/NIRES, which have a finer resolution ($\mathcal{R}$\,$\sim$\,700) than the PRISM data.

\subsection{Single-epoch $M_{\rm BH}$ measurements}
\label{sec:SE_MBH}

We employed the line parameters previously derived to estimate the mass of the black hole powering the accretion in our sample of QSOs. With the aim of exploring the range of possible \Mbh values associated with different calibrations, we took advantage of a number of recipes for each emission line. In particular, we explored calibrations in the form:
\begin{equation}
    \log(M_{\rm BH}) = a + b \, [\log(L) - 44] + c \, [\log(W)-3],
\label{eq:Mbh_vir_eq}
\end{equation}
where $L$ is the continuum or line luminosity in $\rm erg  \, s^{-1}$ and 
$W$ is the line width (either FWHM or $\sigma$) in $\rm km \, s^{-1}$. We adopted the following calibrations (see also Table \ref{tbl:MBH_calibrations}): \citet[VP06]{vestergaard2006determining},  \citet[TN12]{trakhtenbrot2012black}, \citet[B13]{bentz2013low}, \citet[DB20]{db2020}, and \citet[S24]{shen2024sloan} for \Hb; \citet[GH05]{greene2005estimating}, \citet[W15]{woo2015black}, \citet[C23]{cho2023seoul}, and \citet[DB25]{dalla2025estimating} for \Ha; and, lastly, \citet[MD04]{mclure2004cosmological}, \citet[VO09]{vestergaard2009mass}, \citet[S11]{shen2011}, and \citet[S24]{shen2024sloan} for \Mgii. Incidentally, we acknowledge the presence of other prescriptions involving, for instance, the \ion{Fe}{ii}\textsubscript{opt}/\Hb or the \ion{Fe}{ii}\textsubscript{UV}/\Mgii ratios (see, e.g., \citealt{pan2025iron}) as additional terms to account for the effects of the Eddington ratio on the BLR kinematics, which could systematically shift (by $\sim$0.2 dex) the \Mbh estimates towards lower values. However, we expect a general agreement between the \Mbh values, within the (large) systematic uncertainties that similarly affect the prescriptions above and the ones not considered here. It is not trivial to determine the absolute accuracy of SE recipes, because of the several terms contributing to the total uncertainty. While early works reported overall accuracies of the order of 0.5 dex or higher (e.g. \citealt{vestergaard2006determining}), more recent studies lowered it to $\sim$0.4 dex (e.g. \citealt{park2012lick, shen2024sloan}). For the remaining of this work we will assume a systematic uncertainty on SE \Mbh estimates of $\sim$0.4 dex. We discuss the magnitude of this uncertainty in greater detail in Appendix \ref{app:systematics}.

\subsection{Bolometric corrections}
\label{sec:kbol_LBOL}

A bolometric correction (\kbol) offers a convenient way to estimate the bolometric luminosity in sources whose entire SED is not known. From an observational standpoint, bolometric corrections are calculated on (relatively small) samples with an exquisite panchromatic coverage, and then applied to sources with much sparser data (e.g., \citealt{richards2006spectral, lusso2012bolometric, runnoe2012updating, duras2020universal}). Alternatively, they can be derived from theoretical models, whose free parameters are possibly tuned to reproduce observations (e.g., \citealt{marconi2004local, 2010MNRAS.408.1598N, netzer2019bolometric}). Since the latter bolometric corrections are based on AD models, similar to those we employ to derive the accretion parameters (see Section \ref{sec:ad_modelling}), we took advantage of the recipes presented in \citet[NB10]{2010MNRAS.408.1598N} and \citet[N19]{netzer2019bolometric}. These have the form:
\begin{equation}
    \log(L_{\rm bol}) = \alpha + \beta \, \log(\lambda L_{\lambda}),
\label{eq:Lbol_eq}
\end{equation}
where $\lambda L_{\lambda}$ is a monochromatic luminosity at some fixed wavelength free from strong emission lines. Generally, $\lambda$\,=\,1,350~\AA, 2,500~\AA, 3,000~\AA, 5,100~\AA\ are employed, with a preference for 3,000~\AA\ as it is less prone to host-galaxy contamination than 5,100~\AA, and less affected by possible dust (and gas) flux attenuation than the options at bluer wavelengths. For this reason, we used the 3,000-\AA\ luminosity, with $\alpha$\,=\,9.24, 9.80 and $\beta$\,=\,0.81, 0.80 from NB10 and N19, respectively.
As done for the systematic uncertainties on \Mbh, we explore in greater detail the uncertainties related to bolometric corrections in Appendix \ref{app:systematics}. However, here we anticipate that even though some sources of uncertainty can be shared by \Mbh and \Lad calibrations (e.g., the inclination of the line of sight to the AD and BLR), the systematic uncertainties on \Lad are typically much smaller than those on \Mbh ($\sim$\,0.2 dex against $\sim$\,0.4 dex). Therefore, whenever \Lad is combined with \Mbh to estimate \ER, in case of little or no covariance the contribution of \Lad to the total uncertainty becomes negligible. In the following, we will assume an average uncertainty on \Lad estimated via bolometric corrections of 0.2 dex.

\subsection{Accretion-disc modelling}
\label{sec:ad_modelling}
Accretion-disc modelling offers a viable alternative to estimate the accretion parameters underlying the QSO phase (e.g., \citealt{malkan1983ultraviolet, shang2005quasars, capellupo2015active, campitiello2018constrain, cheng2019modelling, campitiello2019black, lai2023characterising, wolf2024accretion, trefoloni2025_ganifs}). This approach requires fairly demanding observational conditions compared to the SE method, such as a wide (possibly simultaneous) coverage of the QSO SED also probing the peak of the UV emission, with a spectral resolution high enough to disentangle continuum windows from emission lines. However, if these requirements can be met, systematic uncertainties are greatly reduced. By providing a large simultaneous coverage of the rest-frame optical and UV SED of QSOs at the epoch of reionisation, \jwst stands out as the only facility suited to pursue this effort on the high-$z$ sources analysed here.

Notably, at these high luminosities, we can neglect the possible presence of galactic contamination at optical wavelengths. Indeed, the flattening of the continuum at optical wavelengths ($\lambda$\,$\gtrsim$\,4,200 \AA) generally associated to the host galaxy (e.g., \citealt{vandenberk2001}) is not observed in any of our targets. For a quantitative comparison, both \citet{jalan2023empirical} and \citet{ren2024prior} propose that sources above $\log(L_{5100\,\AA}/{\rm erg \, s^{-1}})$\,$\gtrsim$\,45.4 have less than 10\% of host galaxy contamination at 5,100~\AA, and our sources have $\log(L_{5100\,\AA}/{\rm erg \, s^{-1}})$\,=\,45.4--46.4.

\begin{figure*}[htbp]
    \centering
    \begin{minipage}{0.55\textwidth}
        \centering
        \includegraphics[width=\linewidth]{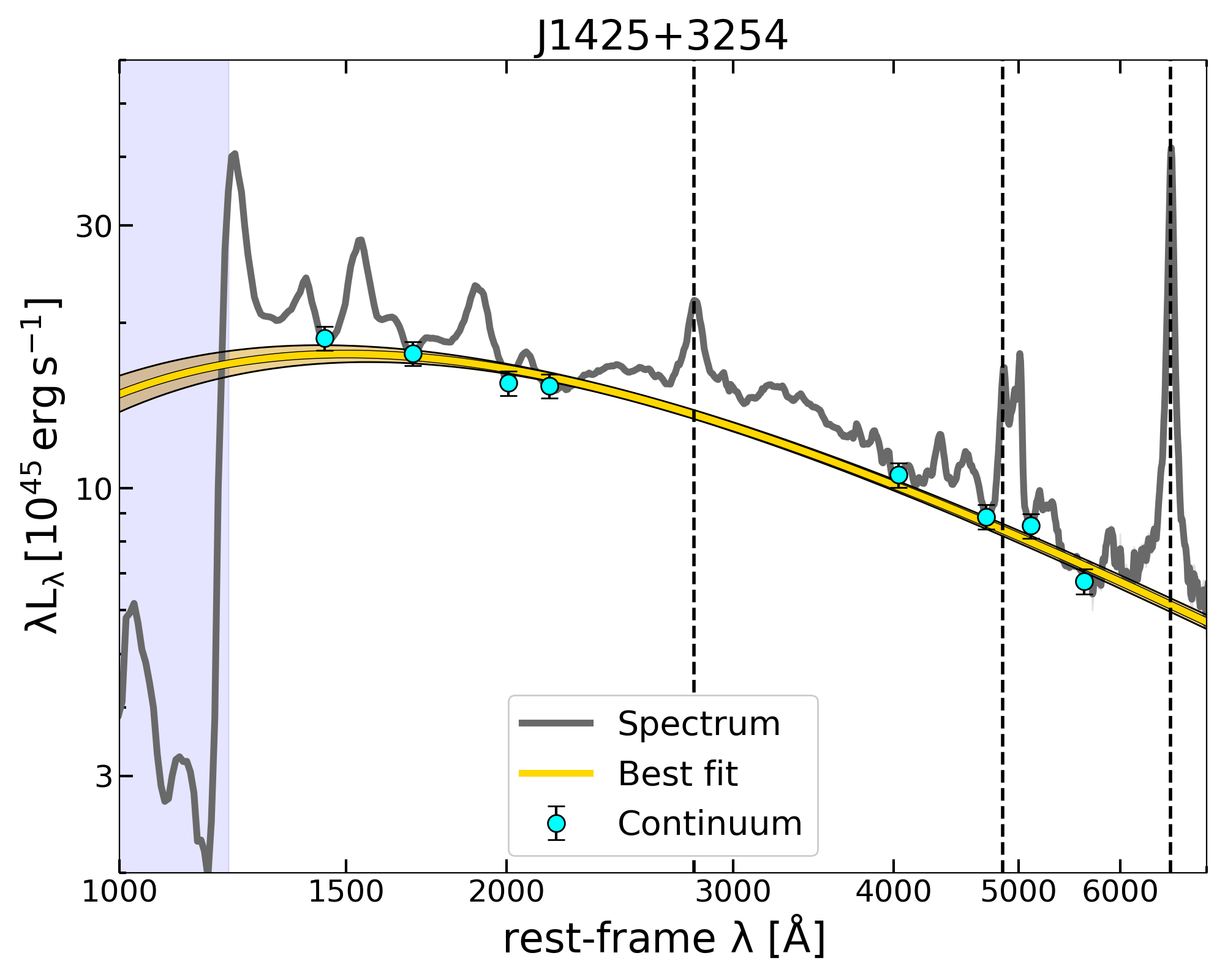}
    \end{minipage}
    \hspace{0.1cm} %
    \begin{minipage}{0.35\textwidth}
        \centering
        \includegraphics[width=\linewidth]{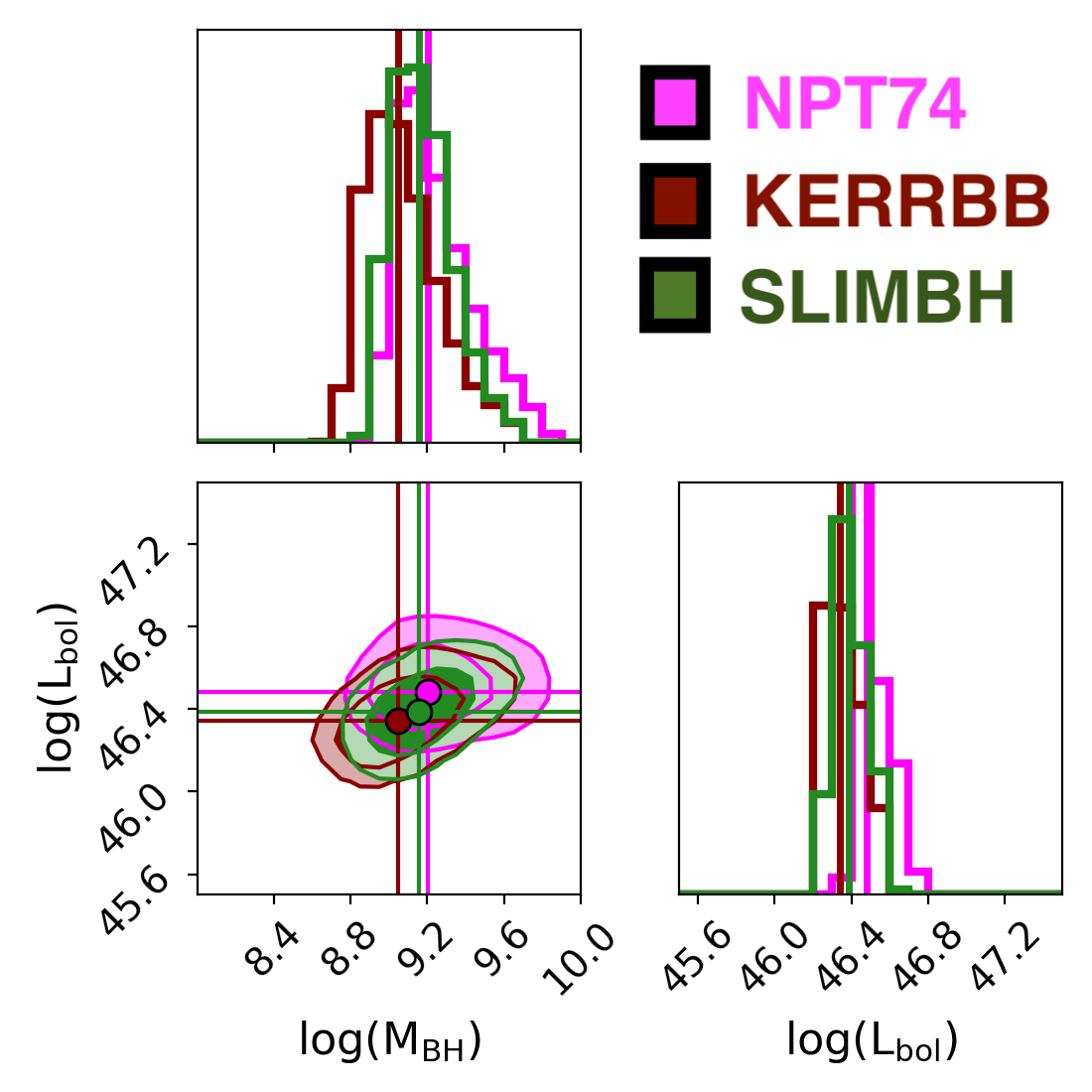}
    \end{minipage}
\caption{\textit{Left:} Accretion-disc modelling of J1425$+$3254. The continuum points employed in the fit are marked as cyan dots. The shaded bands represent the 16$\rm^{th}$--84$\rm^{th}$ percentiles of the distribution of 10,000 randomly extracted best-fit models. The dashed lines mark the broad emission lines (\Mgii, \Hb, \Ha) used to compute \Mbh via SE relations. The shaded area highlights the region bluewards of Ly$\alpha$, affected by IGM absorption and thus excluded from the fit. \textit{Right:} Joint posterior distribution of \Mbh and \Lad.}
\label{fig:ex_ADfit}
\end{figure*}

For the AD modelling, we isolated the continuum emission in small spectral windows ($\sim$\,20--30 \AA, see Fig.~\ref{fig:ex_ADfit}), avoiding strong emission lines as well as the small blue bump between $\sim$\,2,200--4,000 \AA\ \citep{grandi1982}, which consists of the blend of the \ion{Fe}{ii} pseudo-continuum and the Balmer continuum. In particular, the continuum anchor points are centered at 1,475 \AA, 1,690 \AA, 1,810 \AA, 2,150 \AA, 4,035 \AA, 4,685 \AA, 5,130 \AA, 5,655 \AA, 6,020 \AA, and were tuned to avoid broad absorption features and/or noise spikes. Adding a further continuum point close to the Ly$\alpha$ at 1,350 \AA\ does not change the best-fit estimates of \Mbh and \Lad, as the presence of the broad component of this line as well as other fainter lines (\ion{Ne}{V}$\lambda$1240, \ion{Si}{ii}$\lambda$1262, \ion{O}{i}$\lambda$1302, \ion{C}{ii}$\lambda$1335) hampers a proper determination of the continuum. We also introduced a 5\% uncertainty factor in the continuum points to account for the JWST/NIRSpec flux calibration uncertainty.

The AD modelling framework follows closely that already described in \citet{trefoloni2025_ganifs}, to which we refer for further information. Here, we only highlight the main differences with respect to that work. We implemented the general relativistic corrections derived in \citet{novikov1973astrophysics} and \citet[][see their Eq.~15n]{page1974disk} to the geometrically thin and optically thick accretion disc described by \citet{shakura1973black}. We refer to this as the NPT74 model. In order to allow for different mass-to-light conversion efficiencies, we included the dimensionless spin parameter $a^*$ as a free parameter. We also accounted for the disc inclination by introducing a simple wavelength-independent term $f(\theta) = \cos\theta\,(1+d \cos\theta)/(1+d)$, where $d$ is the limb-darkening factor. Here we chose $d$\,=\,2 (e.g., \citealt{netzer2014bolometric, capellupo2015active, netzer2019bolometric}), which is well suited for an electron scattering atmosphere. As complementary approaches, with the aim of including other general relativistic corrections, we also fitted the data employing the KERRBB model, a multi-temperature blackbody model for a thin, steady-state accretion disc around a Kerr black hole (\citealt{li2005multitemperature}), and the SLIMBH model, which accounts instead for the possible thickening of the accretion disc at high accretion rates (\citealt{skadowski2009slim})\footnote{Due to the lack of a systematic X-ray coverage of our sample, we refrained from adopting also models that self-consistently couple the accretion disc and the X-ray corona (e.g., QSOSED; \citealt{kubota2018physical}). For the sake of simplicity, we also avoided exploring more elaborate models such as those coupling AD and BLR (e.g. \citealt{hopkins2024multi}) or with extreme magnetic fields (magnetically-arrested discs; MADs, e.g. \citealt{tchekhovskoy2011efficient}).}. Again, we refer to \citet{trefoloni2025_ganifs} for more details about the implementation of the fitting procedure and the model averaging. The best fit estimates of both \Lad and \Mbh only exhibit minor differences between the different models, with deviations of $\lesssim$\,0.1 dex on average. Even assuming this value as a systematic uncertainty on both parameters our conclusions would be largely unaltered. We also mention that the posterior distributions do not constrain $a^*$ and $\theta$. Hopefully, the future availability of statistically meaningful samples of some tens of high-$z$ QSOs with large spectral coverage as the ones presented here, will allow us to determine -- at least on average -- both of these quantities, therefore providing cosmological simulations of SMBH growth in the early Universe with key ingredients.

As a further test of the reliability of our fitting approach, we also employed the `BADFit' routine (\citealt{lai2023characterising,samuel_lai_2023_7772748}) to independently estimate \Lad and \Mbh, using both the KERRBB and the SLIMBH models within a Bayesian framework. We find a generally good agreement, with an average deviation for \Mbh and \Lad negligible with respect to the typical uncertainties. We show the comparison of the distributions of the residuals between our best-fit models and those derived with `BADFit' in Appendix \ref{app:custom_vs_badfit}.


\section{Results}
\label{sec:results}

We employed the accretion parameters, derived for each source according to different recipes, to investigate the average properties of our sample of QSOs. In particular, we derived the \Mbh, \Lad, and \ER distributions and performed additional tests to take into account the possibility of intrinsic reddening, which could bias our estimates.

\subsection{$M_{\rm BH}$ distribution}
\label{sec:Mbh_distribution}

In Fig.~\ref{fig:mbh_lbol}, we present the comparison between the \Mbh values derived according to the different SE prescriptions employed and the one derived via the AD modelling. When using the DB20 and DB25 calibrations, we explored the results obtained from using both FWHM and $\sigma$ as proxies of the BLR velocity. Also, when taking advantage of the \Mgii line, for which the distinction between the narrow and the broad components is not univocal, we tested the outcomes of employing the broad component derived from a double-Gaussian (one broad and one narrow) fit, and from a single, broad Lorentzian profile.

In general, \Mbh as measured from AD fitting is close to the centre of the SE-based mass distribution. An interesting outcome is the scatter in the SE \Mbh values even for the same line, essentially reflecting the diversity in the calibrations, also depending on the adopted broad-line profile, as in the case of \Mgii. For any given source, the same line yields results spanning $\sim$\,1 order of magnitude. Black-hole masses estimated using the DB20 calibrations for \Hb tend to be higher than the other ones, while the VP06 and TN12 ones are generally closer to the AD value. This is in large part due to the virial factor $\log f$\,=\,0.683 adopted by DB20 (after \citealt{batiste2017recalibration}) for both FWHM and $\sigma$. This value is larger than those adopted in W15 and C23, who chose $\log f$ equal to 0.05 for FWHM- and 0.65 for $\sigma$-based \Mbh. More specifically, the mean deviations in dex  between the AD and SE estimate as a function of the SE calibrations are: $-$0.2 (VP06), $-$0.1 (TN12), $-$0.2 (B13), 0.2 (DB20, with FWHM as line width), and 0.5 (DB20, with $\sigma$). The full list of the deviations can be found in Table \ref{tbl:MBH_calibrations}. 
On the other hand, masses based on \Mgii appear to be more clustered, as testified by the smaller scatter in the deviations (see column 8 in Table \ref{tbl:MBH_calibrations}). Possibly, this is due to the fact that these are not actually independent calibrations, as most of them are basically anchored to the \Hb SE calibrations. \Mgii-based masses obtained adopting a Lorentzian profile appear to provide values slightly more consistent with the other estimates, while the Gaussian ones tend to overestimate the AD mass. In contrast, in J0020$-$3653 the mass obtained with the Gaussian profile is a better match to the AD one. While in the latter case a Gaussian profile might truly be more suitable to reproduce the emission line, this source could also be moderately reddened (see Sect. \ref{sec:extinction}). Therefore, the actual mass could be overestimated by the AD model and underestimated by the SE recipe because of the extinction of the 3,000-\AA\ luminosity. We also note that, since the FWHM of the Lorentzian profile of J2239+0207 is close to the PRISM spectral resolution at the corresponding wavelength ($\sim$\,6,000 km s$^{-1}$ at 2.03 $\mu$m), the actual line width is uncertain, possibly causing the mismatch with the AD estimate.
The \Ha line was not covered in J0313$-$1806, J1007+2115, and J1342+0928, while \Mgii was affected by a strong BAL in J0910$-$0414, therefore no fits were performed for these lines in these sources. We also stress that most of the calibrations derived from samples with RM measurements (VP06, B13, W15, DB20, C23, DB25) were derived for sources in the luminosity range 42\,$\lesssim$\,$\log(L_{5100\,\AA})$\,$\lesssim$\,45, with few of them reaching $L_{5100\AA} \sim 10^{46} \, \rm erg \, s^{-1}$, while most of our sources are at the higher end of (or above) this interval. Therefore, the direct extrapolation of these recipes to our targets could be questionable, due to a different impact of the radiation field, for instance. Reverberation mapping campaigns on bright QSOs will be key to assess their validity (e.g. the 'Black hole mapper' project, \citealt{kollmeier2019sdss}). In what follows, we will assume the TN12 calibrations as our reference ones, since they show the best agreement between AD and SE mass values. While this does not necessarily mean that the TN12 calibration is intrinsically the most accurate one, any other choice would imply a larger average bias between the AD and the SE masses. \\

\subsection{$L_{\rm bol}$ distribution}
\label{sec:Lbol_distribution}

\begin{figure*}[h!]
\centering
\includegraphics[width=\linewidth,clip]{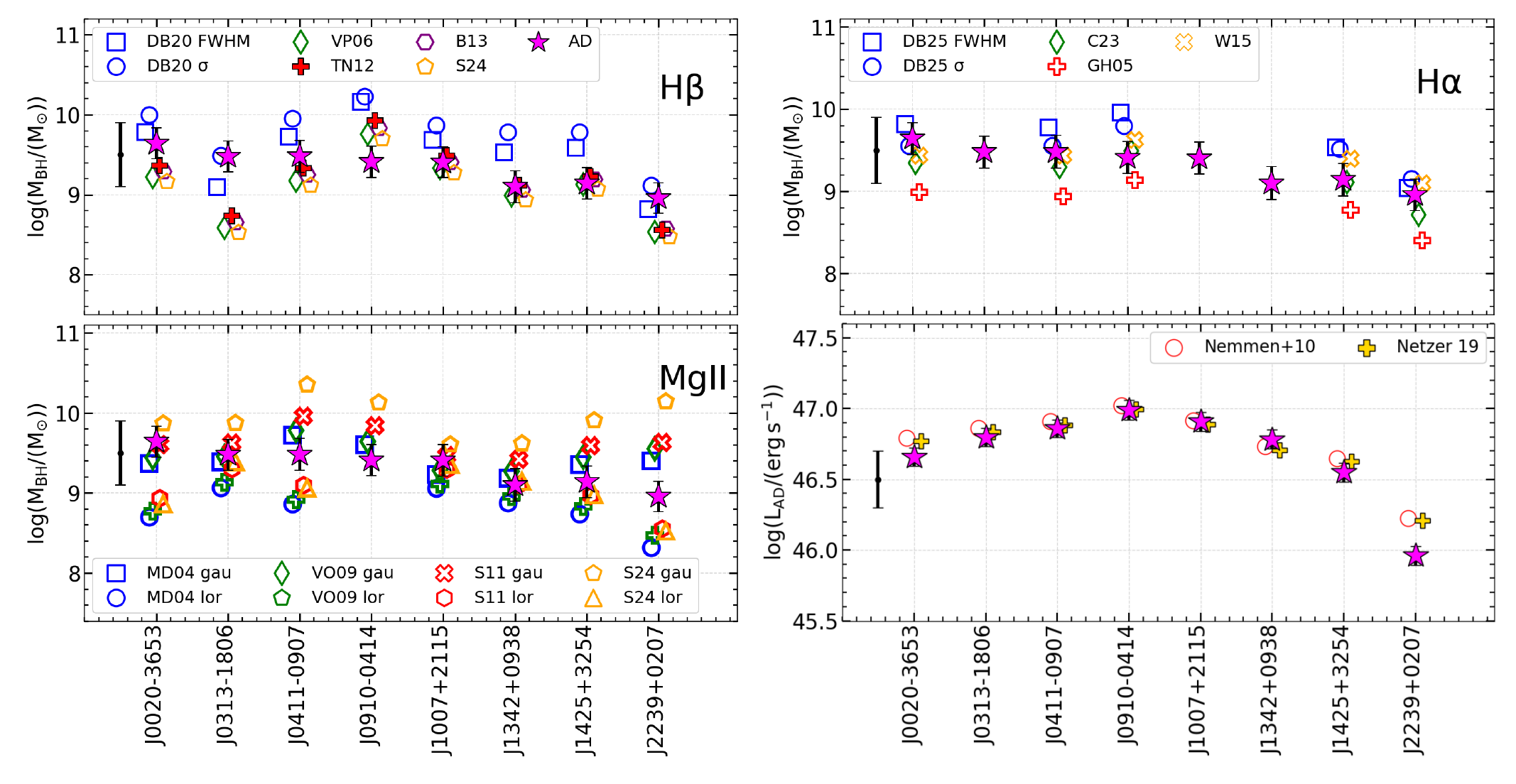}
\caption{Top left, top right, bottom left: comparison between the $M_{\rm BH}$ values estimated from SE calibrations and AD fitting for each QSO. Different colours and symbols represent different prescriptions as listed in Table \ref{tbl:MBH_calibrations}. Only the fiducial TN12 and N19 calibrations are shown with full coloured symbols. In the case of the DB20 and DB25 calibrations, we employed their recipes for both FWHM and $\sigma$ of the line. For the \Mgii calibrations, we used the broad Gaussian (gau) as well as the Lorentzian (lor) line widths. The AD estimate is marked by a magenta star across all panels. Measurement uncertainties are typically smaller than the point size. The errorbar in each panel highlights the 0.4 dex systematic uncertainty (see Section \ref{sec:SE_MBH}). Points are slightly shifted horizontally for visualisation purposes.
Bottom right: comparison between the $L_{\rm bol}$ values estimated from bolometric corrections and AD fitting for each QSO. The systematic uncertainty of 0.2 dex is shown as an errorbar on the left. The worst agreement is found for J2239+0207, for which the discrepancy with respect to the N19 expectation is 0.25 dex.}
\label{fig:mbh_lbol}
\end{figure*}

In the bottom right panel of Fig.~\ref{fig:mbh_lbol} we show the \Lad values derived with our AD modelling and those estimated using the \kbol prescriptions from N10 and N19, specifically their 3,000-\AA\ one. In the latter works, the authors adopted an average inclination of $56^{\circ}$ for their calculations; hence, for the sake of a fair comparison, we re-run our fits after fixing the inclination to that value, since this parameter was basically unconstrained previously. We note, in general, a good agreement, with a mean deviation of, respectively, 0.05 and 0.07 dex for the N19 and the N10 \kbol values, well below the systematic uncertainty. We therefore adopted the N19 calibration as the fiducial one. We note that choosing the \kbol values provided for other wavelengths (e.g., 1,350 \AA, 5,100 \AA) would have produced a slightly lower agreement, yet still within the uncertainty.

\subsection{The $\lambda_{\rm Edd}$ distribution}
\label{sec:er_distribution}

\begin{figure}[h!]
\centering
\includegraphics[width=\linewidth,clip]{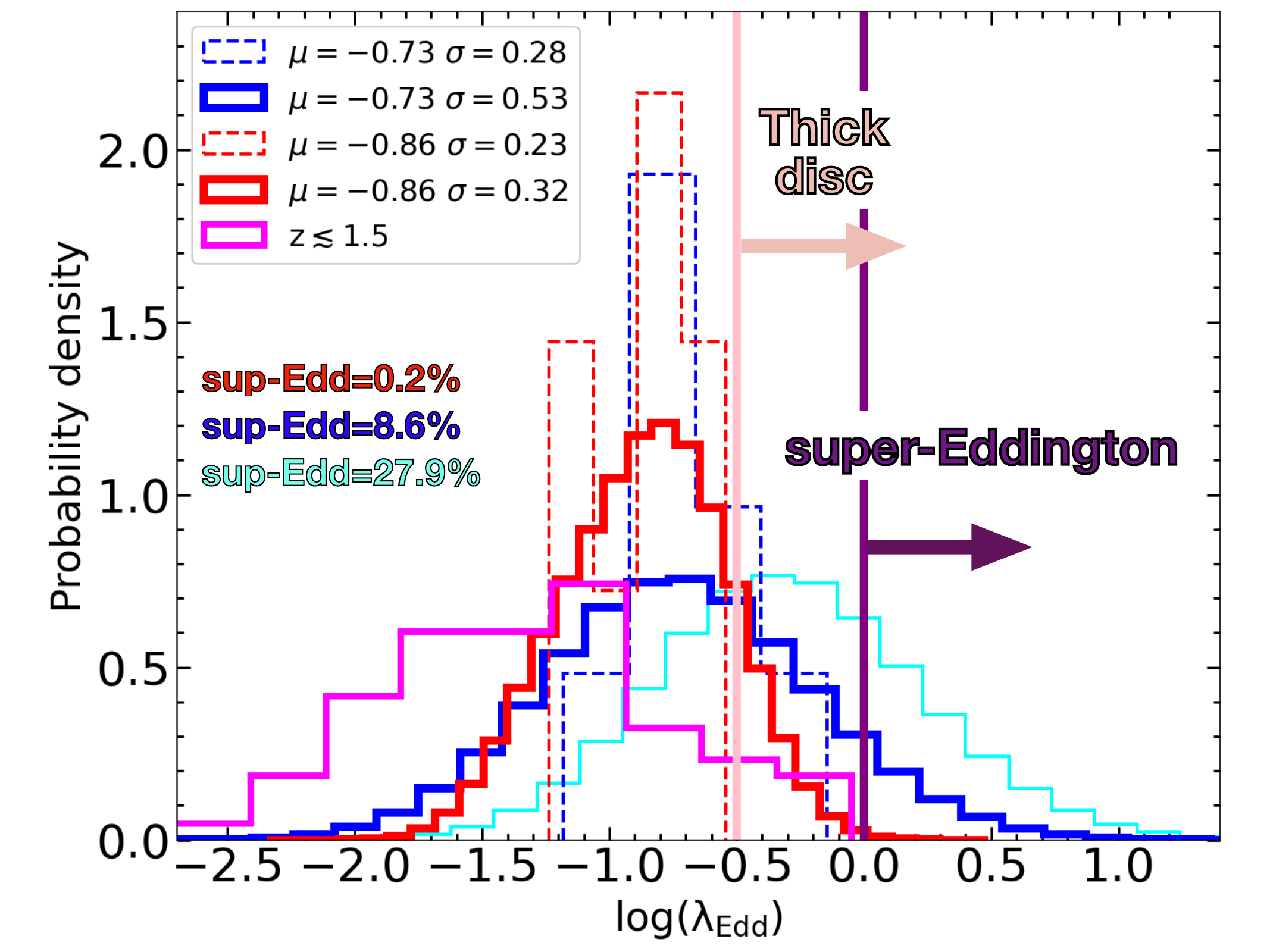}
\caption{Distribution of \ER in our sample. The dashed histograms show the best-fit values distributions, while the solid ones are smoothed for the systematic and statistical uncertainties. Blue and red lines represent the \ER distributions derived from calibrated SE relations and AD modelling, respectively. In magenta we present the total \ER distribution for the joint \citet{capellupo2015active}, \citet{cheng2019modelling}, and \citet{campitiello2020estimating} samples at $z$\,$\lesssim$\,1.5. The cyan histogram highlights the effect of adopting the \citet{richards2006spectral} bolometric correction. We also report the super-Eddington (sup-Edd) fractions of our sample according to different recipes using the same colour-code as the distributions.}
\label{fig:ER_distr}
\end{figure}

We combined our best estimates obtained for \Lad and \Mbh using, respectively, the fiducial \kbol and SE calibrations (we will refer to the latter as the calibrated \Lad and \Mbh), and AD fitting to evaluate \ER in our sources. We show these distributions in Fig.~\ref{fig:ER_distr}. 
We have already shown in Fig.~\ref{fig:mbh_lbol} that if we fix the inclination angle, which is only loosely constrained in the AD fitting process, our \Lad closely follows those derived by both N10 and N19. When providing the \kbol value in N19, the author suggests to decrease the corrections by a factor of $\sim$\,1.4 if the torus has a realistic covering factor of 0.5, as this implies an average inclination smaller than $56^{\circ}$, and by a factor of $\sim$\,2.5 in case of a polar line of sight. We therefore reduced \kbol by 1.4, but it is plausible that all these sources are observed close to face-on inclinations, likely because of selection effects. This possibility has also been suggested by studies on the CO kinematics of QSO host galaxies (e.g., \citealt{carilli2006erratum, wu2007co, ho2007co}), although the disc inclination does not necessarily align with the CO distribution in the host galaxy on much larger scales.

The mean values of the \ER distributions (shown as dashed histograms in Fig. \ref{fig:ER_distr}) are $\rm \langle \log(\lambda_{Edd}) \rangle$\,=\,$-$0.73 and $-$0.86 for the calibrated and AD \ER, with dispersions of 0.28 and 0.23, respectively. In both cases, the typical uncertainties on the individual \ER values are not negligible with respect to the spread of the distributions. Because of this, and with the aim of building a smoother distribution in spite of the small sample size, we folded in the final distributions the uncertainties on both \Mbh and \Lad. We achieved this by considering the systematic uncertainties on the calibrated \Mbh and \Lad, while we employed the uncertainties on the best-fit values in the case of the AD ones. In the latter case, part of the systematic uncertainty deriving from the limited knowledge of the actual AD model was already accounted for through model averaging. To this end, we created two mock samples of 100,000 elements each, by randomising \Mbh and \Lad using a Gaussian distribution whose standard deviation was set to the systematic (best-fit) uncertainty for the calibrated (AD) values. We show the distributions obtained in this way as thick-edged empty histograms in Fig.~\ref{fig:ER_distr}. There, we also show the joint distribution of different works at $z$\,$\lesssim$\,1.5 employing AD modelling to estimate the accretion parameters (\citealt{capellupo2015active, cheng2019modelling, campitiello2020estimating}).

Several interesting features stand out. Although the observed distributions have similar spreads, once smoothed assuming the typical uncertainties as kernel widths, the distribution of the calibrated \ER becomes fairly large, with a standard deviation approaching $\sim$\,0.5 dex. This is mostly driven by the large uncertainty on \Mbh. As we stated in Sec.~\ref{sec:kbol_LBOL}, a systematic uncertainty of 0.2 dex or 0.3 dex on \kbol would not make any significant difference when estimating the spread of the \ER distribution. We also highlight that, albeit after incorporating the systematic uncertainties the calibrated \ER distribution exhibits a super-Eddington tail, the face-value distribution only never reaches the Eddington limit. With the exception of J0313-1806, which has $\lambda_{\rm Edd}\sim 0.7$, thre rest of the sample shows relatively moderate accretion rates at $\lambda_{\rm Edd}< 0.3$.

One of the major consequences of the different spread between the AD and the calibrated \ER distributions is the fraction of super-Eddington sources. The calibrated \ER distribution has a $\sim$\,8.6\% fraction of super-Eddington sources, while the AD one has $\sim$\,0.2\% only. The latter fraction would be even lower if the measurement uncertainty could be reduced, with a better sampling of the AD peak and complementary constraints on the spin or the inclination. While the width of the AD \ER distribution is also partially affected by the $\lambda_{\rm Edd}<1$ cap on the SLIMBH model, this is unlikely to be the driving factor. Indeed, the SLIMBH posteriors largely overlap with those of the other models (see Fig.~\ref{fig:ex_ADfit} and Fig.~\ref{fig:fit_atlas}). Lastly, we note that for $\lambda_{\rm Edd}$\,$\gtrsim$\,0.3 the disc is expected to thicken (e.g., \citealt{abramowicz2013foundations}) and our treatment could not be entirely appropriate, albeit the peak of the best-fit \ER distribution never reaches such a value.

This resonates with the findings of other works taking advantage of AD modelling (\citealt{capellupo2015active, cheng2019modelling, campitiello2020estimating}) to estimate \Mbh and \ER, albeit with slightly different models.\footnote{Also \citet{calderone2013black} estimated \Mbh on a sample of 23 radio-loud Seyfert 1 galaxies at $z$\,=\,0.1--0.8, using a method that combined broad lines and AD modelling. Although slightly different from the others mentioned here, in that case too the authors found that all their sources are accreting at $\lambda_{\rm Edd}$\,$\lesssim$\,0.1.} There is not even one source out of the 73 in those three samples that is implied to accrete at super-Eddington rate. Here, we also caution that by employing a constant bolometric correction, such as the widely employed \kbol\,=\,5.15 at 3,000 \AA\ from \citet{richards2006spectral}, most of our sources would result in a near- or super-Eddington accretion state. In order to give a tangible example of this, we also included in Fig. \ref{fig:ER_distr} the \ER recalculated using the above \kbol. The limitations of such an approach are discussed in Appendix~\ref{app:systematics} (see also \citealt{trakhtenbrot2012black}). In brief, a constant bolometric correction tends to underestimate the actual correction at low luminosity and overestimate it at the high end (see also Fig. 2 in \citealt{netzer2019bolometric}). If such luminosity-independent \kbol were adopted, in combination with the TN12 SE calibration, the peak of the \ER distribution would shift upwards by $\sim$\,0.4 dex, and the fraction of super-Eddington sources would increase to $\sim$\,28\%.

\subsection{The effect of extinction}
\label{sec:extinction}

The presence of gas and dust along the line of sight within the BLR and/or the host galaxy could bias our estimates of \Mbh and \Lad. In this section, we present several arguments against this effect playing a major role in the observational appearance of our targets. 

Concerning the \Mbh and \Lad calibrations, since both quantities scale with some monochromatic luminosity (or, equivalently, some line luminosity), they would be underestimated with respect to the actual value in the case of flux attenuation. For the AD modelling instead, assuming that differential reddening (i.e., a non greybody-like extinction curve) is present, the observed SED would be dimmer and redder (appearing as a colder AD) than the intrinsic one. This would produce a lower \Lad and a higher \Mbh as a result of the AD fitting. Because of these combined effects, a significant amount of dust reddening would produce a systematic trend in the deviations between the \Mbh values calculated according to SE and AD recipes. In particular, as the magnitude of extinction increases at smaller wavelengths, the difference between the SE and the AD estimates ($\Delta \log M= \log(M_{SE})-\log(M_{AD})$) should become more negative going from \Ha to \Hb and then to \Mgii. We do not observe this trend.

Although we do not expect our QSO sample to be significantly affected by dust, as it consists of some of the most luminous quasars shining in the early Universe, we aimed at testing this hypothesis. Emission line-based methods, such as the Balmer decrement, would not provide a robust estimate of reddening. On the one hand, the narrow-line Balmer ratio only provides a lower limit to the total extinction along the line of sight to the nucleus. Additionally, the narrow lines are not easy to disentangle in the case of sources as bright as ours (and with the spectral resolution offered by the NIRSpec PRISM in half of the sample). On the other hand, the broad-line Balmer ratio is not directly interpretable within the case B recombination framework (e.g. \citealt{osterbrock2006astrophysics}). The extreme physical conditions characterising the BLR are such that collisional, optical-depth, and radiative-transfer effects become important, therefore altering the simple case B recombination ratio (see, e.g., \citealt{korista2004optical}, and references therein).

We thus estimated the amount of dust extinction along the line of sight by reddening a template made of bright blue QSOs, which we assume not to be affected by any dust reddening (e.g., \citealt{glikman2012first, jiang2013anomalously, krogager2016extended, krogager2016quasar}). In particular, we took advantage of the spectral template derived in \citet{selsing2016x}, made of seven of the most luminous blue QSOs at 1\,$<$\,$z$\,$<$\,2 observed with VLT/XSHOOTER. This procedure is analogous to assuming that the modal colours of the distribution these objects are drawn from are dust free. A Small Magellanic Cloud (SMC) extinction curve has been found to well reproduce the outliers in the SDSS QSO colour distribution (e.g., \citealt{hopkins2004dust, krawczyk2015mining}), hence we adopted this curve (referred to as `G24\_SMCAvg') from the Python package `dust\_extinction' (\citealt{gordon2024dust_extinction}) to redden the adopted template, 
choosing a total-to-selective ratio of $R_V$\,=\,3.0 \citep{gordon2024expanded}. We performed the reddening in steps of 0.025 in $E(B-V)$, measuring the spectral slope between the rest-frame 2,200-\AA\ and 5,500-\AA\ fluxes, and then estimated the $E(B-V)$ for each of our sources by means of linear interpolation. The extinction in three out of eight sources turned out to be negative, implying a bluer continuum than the one of the adopted template. The mean $E(B-V)$ derived is 0.02 (0.03 neglecting negative values). The largest $E(B-V)$ is 0.08, found for J0020$-$3653. Adopting a different extinction curve (e.g., \citealt{gallerani2010extinction}) would not produce any appreciable difference.

We took a step further by directly introducing dust reddening into our modelling framework. We achieved this by taking advantage of the QSO compilation assembled in \citet{krawczyk2015mining}, who explored the distribution of reddening in $\sim$\,35,000 sources between 0\,$<$\,$z$\,$<$\,5.3, also distinguishing between BAL and non-BAL sources.\footnote{The bulk of their distribution ($\sim$\,95\%) is at redshifts below 3. We assumed that the $E(B-V)$ distributions for BAL and non-BAL QSOs can be extended to the redshifts sampled by our work.} In brief, we selected all the sources in their sample ($\sim$\,3,100) within $\pm$\,1$\sigma$ from our average $\log(L_{3000\,\AA})$, and computed the $E(B-V)$ distributions for both the BAL and non-BAL objects, as BAL QSOs generally display redder colours (see, e.g., \citealt{yamamoto1999continuum}). We then fitted the $E(B-V)$ distributions, only including positive values using a half-Gaussian function.\footnote{Negative $E(B-V)$ would imply sources with a bluer continuum than the one assumed to be unextinguished. Albeit such sources are interesting for understanding the shape of the unobscured continuum, including them would make the average $E(B-V)$ smaller. Therefore, our approach to avoid such objects gives even tighter constraints on the possible intrinsic reddening.} We used these distributions as priors for the $E(B-V)$ values entering the model likelihood, using as starting guess the $E(B-V)$ values derived at the template reddening stage, and setting the maximum $E(B-V)$ to 0.3. As done for the template reddening, we reddened the disc SED using the `G24\_SMCAvg' extinction curve. Overall, the inclusion of reddening within our modelling only introduces minor changes in the average \Mbh and \Lad distributions, which shift, respectively, by $-$0.11 dex and 0.19 dex. As a result, the average \ER increases from 0.14 to $\sim$\,0.28, yet it remains safely below the Eddington limit. Assuming that the discrepancy with respect to the template is not due to a genuine shift of the peak due to a different disc temperature, but rather to reddening, if we exclude the most reddened sources with E(B-V)$\geq$0.04 (J0020-3653, J0411-0907, J2239+0207), the average $\langle \lambda_{\rm Edd}\rangle$ shifts to -0.77 while the dispersion shrinks to 0.26. Since the inclusion of dust reddening acts in the direction of increasing \ER, our conclusion of predominantly sub-Eddington accretion should be regarded as conservative.

\section{Discussion}
\label{sec:discussion}

The presence of SMBHs when the Universe was less than 1 Gyr old is apparently puzzling. It is hard to reconcile the observed black hole masses with the standard accretion picture, unless we postulate massive seeds to start accreting onto or constant super- or even hyper-Eddington accretion. In this framework, obtaining more accurate masses and self-consistently coupling them to the emitted luminosity is of paramount importance.

Despite being the most widely employed, SE-based \Mbh determinations mostly suffer from a large systematic uncertainty inherent to these calibrations. As we explain in detail in Appendix \ref{app:systematics}, numerous effects concur to make these estimates fairly problematic: examples of these are the uncertainty on the virial factor $f$, the scatter in the $R$--$L$ relation, the possible presence of host-galaxy contamination, and the differential line/continuum variability (see, e.g., \citealt{denney2009systematic, park2012lick}), although the latter two should be minimal in case of luminous QSOs. Moreover, while both the FWHM and $\sigma$ are widely employed to characterise the width of virial lines, they are not interchangeable (see, e.g., Section 1.2 in \citealt{db2020}), and it is not possible to fully encapsulate the line properties in just one parameter. Adopting simple analytic profiles for the emission line, such as a Gaussian or a Lorentzian shape, instead of the direct use of the line profile itself, can lead to several systematics (e.g., \citealt{denney2009systematic}). Even in the case of the benchmark \Hb line, it can be hard to disentangle contaminating agents, such as the optical \Feii, or outflow components, from the virial line profile (see, e.g., \citealt{joly1988contribution}). In some cases (e.g. Mrk 509, PDS 456), the virialisation itself of the BLR clouds has been questioned, as dynamical modelling of interferometric data revealed outflow-dominated BLRs (\citealt{amorim2024size}). In this context, the ability of accretion-disc modelling to simultaneously constrain \Mbh and \Lad becomes crucial, especially at high redshift, where RM is unfeasible.

Another viable way to jointly estimate \Mbh and \Lad in AGN and QSOs is offered by AD modelling (e.g., \citealt{koratkar1999ultraviolet}). This method is not as straightforward and widely applicable as the SE one, since it relies on a broad spectroscopic coverage of the Big Blue Bump (\citealt{grandi1982}) that can hardly be obtained with ground-based facilities. Indeed, most of the works employing this approach resorted to VLT/XSHOOTER spectroscopy (e.g., \citealt{capellupo2015active, lai2023characterising, wolf2024accretion}), or were instead forced to combine several non-simultaneous, multi-wavelength measurements (e.g., \citealt{campitiello2020estimating}). Additional complications are the effects of the host-galaxy contamination, especially in the case of faint AGNs, and of intrinsic reddening, which can both distort the global SED. Also, in the case of $z$\,$\gtrsim$\,0.5 sources, the intervening IGM absorbs a large amount of the flux bluewards of Ly$\alpha$ (e.g., \citealt{moller1990lyman}), making it virtually impossible to access the spectral information within that range. Crucially, for the bulk of the QSO population the peak of the emission is expected to lie at wavelengths close to or shorter than the Ly$\alpha$ (e.g., \citealt{cai2023universal}). This notwithstanding, AD modelling allows a self-consistent determination of the accretion parameters, and it is subject to lower systematic uncertainties. As \Lad and \Mbh are simultaneously derived, this approach provides a more robust treatment of the statistic covariance between \Lad and \Mbh, which is key to estimate the \ER distribution. Moreover, AD estimates of \Mbh overcome at least two setbacks of the emission line-based ones, bypassing both the ambiguity of the line profile (i.e., the fact that the component associated with the BLR emission is not always trivial to determine) and the choice of the calibration, which is itself another source of arbitrariness. As RM is unfeasible at high redshift (if anything due to the cosmological time dilation), and the extrapolation of local SE recipes presents several critical aspects, AD modelling is arguably the most sensible method to determine the accretion parameters of QSOs in the early Universe.

In this work, we showcased how the new capabilities delivered by \jwst/NIRSpec offer a possible route for inferring \Mbh in high-redshift ($z$\,$\gtrsim$\,5) QSOs via AD modelling. Besides demonstrating the feasibility of this approach with these new datasets, the main result of our work is the lack of evidence for super-Eddington accretion in these sources. Although our targets constitute a somewhat heterogeneous sample, being the collection of observations with different science goals, their rest-frame optical and UV properties closely follow those of larger samples of optically-selected blue QSOs at both low and high redshift (see also Appendix \ref{app:composite_spectra} for a more quantitative comparison). This argues in favour of our sources being representative of the typical accretion properties of bright blue QSOs. Super-Eddington accretion in high-$z$ QSOs might well be possible, but in systems observationally different from the ones studied here, such as red, dusty (possibly post-merger) systems where large amount of gas can be available to fuel the accretion process (see, e.g., \citealt{kawakatu2007anticorrelation}). Yet, such sources would be hard to disentangle in optical surveys from sub-Eddington accreting QSOs, appearing as reddened due to the line of sight crossing the dusty torus.

Our results align with those derived from other samples where \Mbh and \Lad were self-consistently estimated via AD modelling (\citealt{capellupo2015active, cheng2019modelling, campitiello2020estimating}). Also other works focusing on AD modelling of individual high-redshift QSOs, such as J2157$-$3602 at $z$\,=\,4.7 (\citealt{lai2023characterising}), J0100+2802 -- the most luminous quasar known -- at $z$\,=\,6.3 (\citealt{wolf2024accretion}), ULAS J1120+0641 ($z$\,=\,7.1), DELS J0038$-$1527 ($z$\,=\,7.0), and ULAS J1342+0928 ($z$\,=\,7.5), provided evidence for sub-Eddington accretion (see \citealt{campitiello2019black} and \citealt{trefoloni2025_ganifs}. In this picture, the large fraction of super-Eddington sources, especially found in bright QSOs (e.g, \citealt{mazzucchelli2023xqr, yang2023spectroscopic}), possibly stems from the combination of an overestimation of \Lad due to a fixed bolometric correction and the systematic errors on \Mbh caused by the adoption of local SE calibrations. This suggests that, if either \Lad in QSOs samples were homogeneously re-evaluated using luminosity-dependent \kbol values, or also \Mbh were estimated via AD modelling, a large fraction of the super-Eddington high-$z$ QSOs would vanish. According to our sample, the actual incidence of blue QSOs above the Eddington limit is likely to be $\sim$\,0.2\%, or even lower. Intriguingly, the lack of widespread super-Eddington accretion, among a large sample of optically-selected SDSS QSOs ($\gtrsim$\,100,000 objects), is also suggested by Risaliti, Salvati \& Trefoloni (in prep.), who explain the Baldwin effect (\citealt{baldwin1977}) in terms of an optically-thick, geometrically-thin accretion disc, assuming that most of optically thin broad lines are good proxies of \Lad. As \ER is one of the free parameters in their model, the small dispersion found in the scaling relations across several lines points to an intrinsically narrow ($\lesssim$\,0.2 dex) \ER distribution peaking at $\log(\lambda_{\rm Edd})$\,$\sim$\,$-0.9$. While the central value is in keeping with the results presented here, the narrow width of their \ER distribution, also due to the larger and more representative sample, implies a remarkably low fraction of super-Eddington sources.

In general, it is also worth noting that while the Eddington ratio is a convenient proxy of the accretion state, it only represents a rather crude approximation of the radiative limit. More realistically, an Eddington ratio of $\sim$\,0.2--0.3 could be already enough to alter the structure of the 
accretion disc, making it geometrically thicker (e.g., \citealt{abramowicz1988slim}). In such an accretion state, only part of the entropy generated locally by dissipative processes is released in the form of radiation, while the remainder would be advected by the infalling gas. In this regime, the mass-to-light conversion efficiency is expected to decrease as a fraction of the photons generated in the disc cannot escape, being trapped in the radial flow (e.g., \citealt{begelman1982thick}). More realistic modelling studies, taking into account the time delay between energy generation deep inside the disc and energy release at the surface, showed that $L_{\rm AD}$\,$\sim$\,$L_{\rm Edd}$ even for supercritical mass accretion rates (e.g., \citealt{ohsuga2005supercritical}). This provides a natural framework in which rapid early SMBH growth can occur even without invoking sustained super-Eddington luminosities. 

Another way to regulate the accretion rate and prevent it from reaching values close to or above the Eddington limit is through the launch of powerful outflows (\citealt{slone2012effects, laor2014line}). At $\lambda_{\rm Edd}$\,$\sim$\,0.2--0.3 the radiation pressure on free electrons can already be enough to sustain nuclear winds with sizeable mass accretion and mass outflow rates (e.g., \citealt{zubovas2013bal, nardini2015black, king2015powerful, nardini2019towards}). The propagation of these winds through the host galaxies is then expected to result in the injection of energy and momentum in the inter-stellar medium, yielding the potential to ultimately deliver an early feedback (e.g., \citealt{sijacki2007unified, dimatteo2008direct, harrison2018agn}). Such outflow-driven regulation may be particularly important at high redshift, where recent JWST observations have uncovered overmassive SMBHs accreting at extremely low \ER (e.g. \citealt{juodvzbalis2024dormant}). These findings are consistent with models invoking short bursts of super-Eddington growth accompanied by massive gas expulsion and early feedback, followed by extended dormant phases.

\section{Conclusions}
\label{sec:conclusions}
In this paper, we analysed \jwst observations of a sample of luminous quasars at the epoch of reionisation, employing different recipes to estimate their accretion parameters (i.e., \Mbh, \Lad, \ER).  
Our main findings are listed below:

\begin{itemize}

    \item We adopted a semi-parametric approach to extract the line profile, in order to estimate \Mbh based on several emission lines (\Ha, \Hb, \Mgii) and \Lad based on monochromatic continuum luminosities. The degree of agreement among the various estimates depends on the prescriptions employed and on the assumed proxy for the BLR velocity (FWHM, $\sigma$). Most estimates are consistent with each other, if we include the systematic uncertainty on each \Mbh value (of the order of 0.4 dex). 
    
    \item We employed several optically-thick accretion-disc models to self-consistently estimate \Mbh and \Lad. Although the spin and the inclination angle remained largely unconstrained, all the models produced consistent measurements of both \Mbh and \Lad. Crucially, the average uncertainties on the best-fit \Mbh and \Lad measurements are respectively of the order of 0.2 dex and 0.1 dex, significantly smaller than those associated with single-epoch calibrations and bolometric corrections. Both the best-fit values and the magnitude of the uncertainties are consistent with those independently estimated via the `BADFit' routine. 
    
    \item The \ER distribution derived from AD modelling has a mean value of $\rm \langle\log(\lambda_{Edd})\rangle$\,=\,$-$0.86, with a relatively small standard deviation of 0.23 dex (0.32 dex after smoothing the distribution according to the best fit uncertainties on both \Mbh and \Lad). Assuming that the described sample of high-$z$ QSOs is representative of the entire bright blue QSO population, we derive a fraction of systems accreting above the Eddington limit of $\sim$\,0.2\%, challenging the paradigm of widespread super-Eddington accretion in quasars at the epoch of reionisation.

\end{itemize}

The analysis presented here on a sample of quasars at the epoch of reionisation showcased, once again, the extraordinary capabilities of \jwst. While the high-resolution spectroscopy allowed us to compute the line kinematics and estimate \Mbh according to the usual recipes, an equally crucial amount of information can be gathered from the overall SED captured in both the fixed slit and the low-resolution modes. In a wider perspective, this work showed unambiguously how data collected with \jwst/NIRSpec PRISM and fixed slit could be employed to robustly constrain the black-hole masses in bright QSOs at high redshift, therefore offering an alternative to the widely employed, yet highly uncertain, single-epoch calibrations. Large spectroscopic surveys adopting this observational setup, also favoured by the shorter exposures compared to the medium- and high-resolution modes, could pave the way towards a more thorough understanding of the SMBHs populating the primordial Universe.

\begin{acknowledgements}
This work is based on observations made with the NASA/ESA/CSA James Webb Space Telescope. The data were obtained from the Mikulski Archive for Space Telescopes at the Space Telescope Science Institute, which is operated by the Association of Universities for Research in Astronomy, Inc., under NASA contract NAS 5-03127 for JWST. BT, SC acknowledge support by European Union’s HE ERC Starting Grant No. 101040227 - WINGS. Views and opinions expressed are however those of the authors only and do not necessarily reflect those of the European Union or the European Research Council Executive Agency. Neither the European Union nor the granting authority can be held responsible for them. MS acknowledges support through the European Space Agency (ESA) Research Fellowship Programme in Space Science. AS acknowledges support by the national doctoral scholarship from the Agencia Nacional de Investigaci\'on y Desarrollo (ANID), folio de postulaci\'on 21221788.
\end{acknowledgements}

\bibliographystyle{aa} 
\bibliography{bibl}

\begin{appendix}

\section{Discussion about the systematics on bolometric corrections and SE calibrations}
\label{app:systematics}
Here we discuss in greater detail the systematic uncertainties affecting both the SE \Mbh estimates and the bolometric corrections for \Lad.
We start with the former. While SE calibrations provide an extremely convenient way to estimate \Mbh, it is also sensible to understand their limitations and the associated systematic uncertainties. The largest reverberation mapping campaigns to estimate the time lags between continuum and line emission have targeted the \Hb transition (e.g., \citealt{peterson2004central, bentz2015agn}) for both physical and technical reasons (see, e.g., \citealt{dalla2025estimating} for a discussion on why \Hb is preferred over \Ha). Since widely employed \Ha and \Mgii calibrations are based on the \Hb ones (e.g., \citealt{mclure2004cosmological, greene2005estimating, vestergaard2009mass, shen2011}), their systematic uncertainties should be at least as large. For this reason, we limit the discussion of systematic uncertainties to those concerning the \Hb calibrations, keeping in mind that all the issues discussed for this line largely affect the others too. 

The virial SE calibrations are derived by requiring the virial product ($\mu = V^2 R /G$, where $V$ and $R$ are, respectively, the BLR rotational velocity and radius) to be the same as that measured from RM campaigns. Here, $V$ is derived from the line profile, while the $R$--$L$ relation is used to estimate the BLR radius. 

A crucial uncertainty is that on the virial factor $f$, which is generally tuned to reproduce the $M_{\rm BH}$--$M_{\sigma}$ relation computed on local inactive galaxies. Early works reported a spread by a factor of $\sim$\,2.9 ($\sim$\,0.46 dex; \citealt{onken2004supermassive}). Alternatively, an upper limit on the uncertainty on $f$ of 0.43 dex has also been derived from the scatter of the $M_{\rm BH}$--$M_{\sigma}$ relation on a sample of AGN, assuming this to be representative of the whole AGN population as done in \citet{woo2015black}. More recently, \citet{williams2018lick} lowered this uncertainty to $\sigma_{\log f}$\,=\,0.14\,$\pm$\,0.10, based on 16 objects with RM data used to derive dynamical masses. Yet, the latest SDSS RM campaign found a dispersion in $f$ still of the order of 0.3 dex (\citealt{shen2024sloan}). This uncertainty directly propagates into the uncertainty on RM masses. Furthermore, other sources of uncertainty affect the \Mbh calibrations, such as variability (e.g., \citealt{denney2009systematic}), the intrinsic scatter in the $R$--$L$ relation (e.g., \citealt{bentz2009black}), and the contribution of non-virial components to the line profile (e.g., \citealt{denney2012outflows,coatman2017}). All these effects are summarised in detail, for instance, in Section 1 of \citet{park2012lick}. However, the share of uncertainty introduced by these further systematics is smaller than the one on the virial factor. Lastly, we note that statistical measurement errors on both $L$ and $W$, which, for spectra with reasonably good S/N, are well below 0.1 dex, remain a minor source of uncertainty. In light of these considerations, in this work we assumed 0.4 dex a fiducial systematic uncertainty on SE calibrations.

In addition to effects that produce a symmetric spread in the SE estimates, there are also effects that might introduce systematic biases. In this context, the extrapolation of the SE calibrations to the quasar luminosity regime, well above the range over which they were calibrated, should be treated with caution. With regard to this, as already noted in Section \ref{sec:Mbh_distribution}, our sample lies at luminosities higher than those of the low-redshift sources used to derive SE calibrations ($\log(L_{5100\AA}/(\rm erg \, s^{-1}))\lesssim 45.0$). Above this range, the $R-L$ relation may flatten (\citealt{amorim2024size}), therefore undermining the adoption of such recipes in this regime. A further reason for caution is that SE recipes are calibrated on a subsample of sources exhibiting sufficiently large variability to yield robust time-lag measurements. For instance, significant lags ($>2\sigma$) were detected in only 125 out of the 714 sources with observations covering the \ion{Mg}{ii} sample in \citealt{shen2024sloan} and an even smaller fraction (25/453) was recently reported in \citealt{mcdougall2025ozdes}. While this may also reflect a non-optimal time sampling and S/N limitations, it remains unclear how representative this subsample is of the global QSO population. For similar reasons, the adoption of these recipes to estimate \Mbh in AGNs with BLR kinematics and physical conditions possibly extremely different from local AGN, such as the population of broad line AGN discovered by JWST (see e.g. \citealt{juodvzbalis2024jades, maiolino2025jwst}), or the even more remarkable sub-population of little red dots (LRDs; \citealt{matthee2024little}), is, at least, questionable.

We now focus on the sources of systematic uncertainties on bolometric corrections. Empirical bolometric corrections are more robust, being based on observations, and intrinsically incorporate the parameter space explored by the joint \Mbh and \Lad distributions. Yet, they present several setbacks, such as the need to disentangle the AGN and the galaxy emission, the contamination from emission lines that needs to be corrected for, and the poor coverage of the extreme UV (see, e.g., \citealt{runnoe2012updating}). An additional issue is that different authors employ different integration limits to define their bolometric corrections. For instance, \citet{elvis1994atlas} and \citet{richards2006spectral} integrated their entire average SEDs, from the X-rays to the far-infrared. Instead, \citet{marconi2004local} and \citet{runnoe2012updating} limited their integration up to 1 $\mu$m, in order to avoid `double-counting' of the disc photons absorbed and re-emitted by the dusty torus in the infrared. Reiterating the argument already made in \citet{trakhtenbrot2012black}, both observations and theory provide evidence against a constant bolometric correction. Such corrections are indeed expected (and observed) to decrease with increasing optical luminosity, while the opposite is true for the X-ray bolometric correction (e.g., \citealt{netzer2019bolometric}). Therefore, adopting a constant bolometric correction tends to overestimate the actual \Lad in bright AGN (above $L_{5100\,\AA}$\,$\sim$\,10$^{45}$ erg s$^{-1}$). 

The anisotropy of the disc emission carries a large share of the uncertainty on \kbol. Since the main source of luminosity in the rest-frame optical/UV is the accretion disc, whose orientation with respect to the line of sight is generally unknown, using an average inclination introduces a systematic uncertainty on the correction. NB10 estimated the fractional uncertainty on \kbol, showing that the isotropic assumption can lead to an under-/over-estimation of the intrinsic luminosity by factors larger than 30\%. Also the lack of a precise knowledge of the other accretion parameters, such as $a^*$ and \Mbh, contributes to the uncertainty on \kbol. For instance, the \kbol value presented in N19, although calculated with a fixed line of sight of $\theta = 56^{\circ}$, still presented a spread as high as 0.35 dex, as a result of the marginalization over the unknown parameters. This value is somewhat smaller in empirical bolometric corrections, which have uncertainties of the order of 30--40\% ($\sim$\,0.15 dex; \citealt{runnoe2012updating}). This is likely due to the fact that the line of sight does not span a large interval of inclination angles in broad-line AGN, implying that the whole parameter space is not practically accessible. 

Lastly, we mention that some systematic uncertainties on both \Mbh and \Lad derive from the lack of knowledge of the same physical quantity. For instance, the inclination of the line of sight affects both parameters. However, if we are interested in quantifying the resulting spread of the \ER distribution, the total systematic uncertainty on \ER can hardly be lower than that on \Mbh alone. A theoretical effort to self-consistently model the accretion disc and BLR will be key to evaluate the extent of the effects of the possible covariance between the systematic uncertainties.

\section{Comparison between BADFit and our accretion parameters}
\label{app:custom_vs_badfit}
\begin{figure*}[h!]
\centering
\begin{subfigure}[t]{0.48\linewidth}
    \centering
    \includegraphics[width=\linewidth,clip]{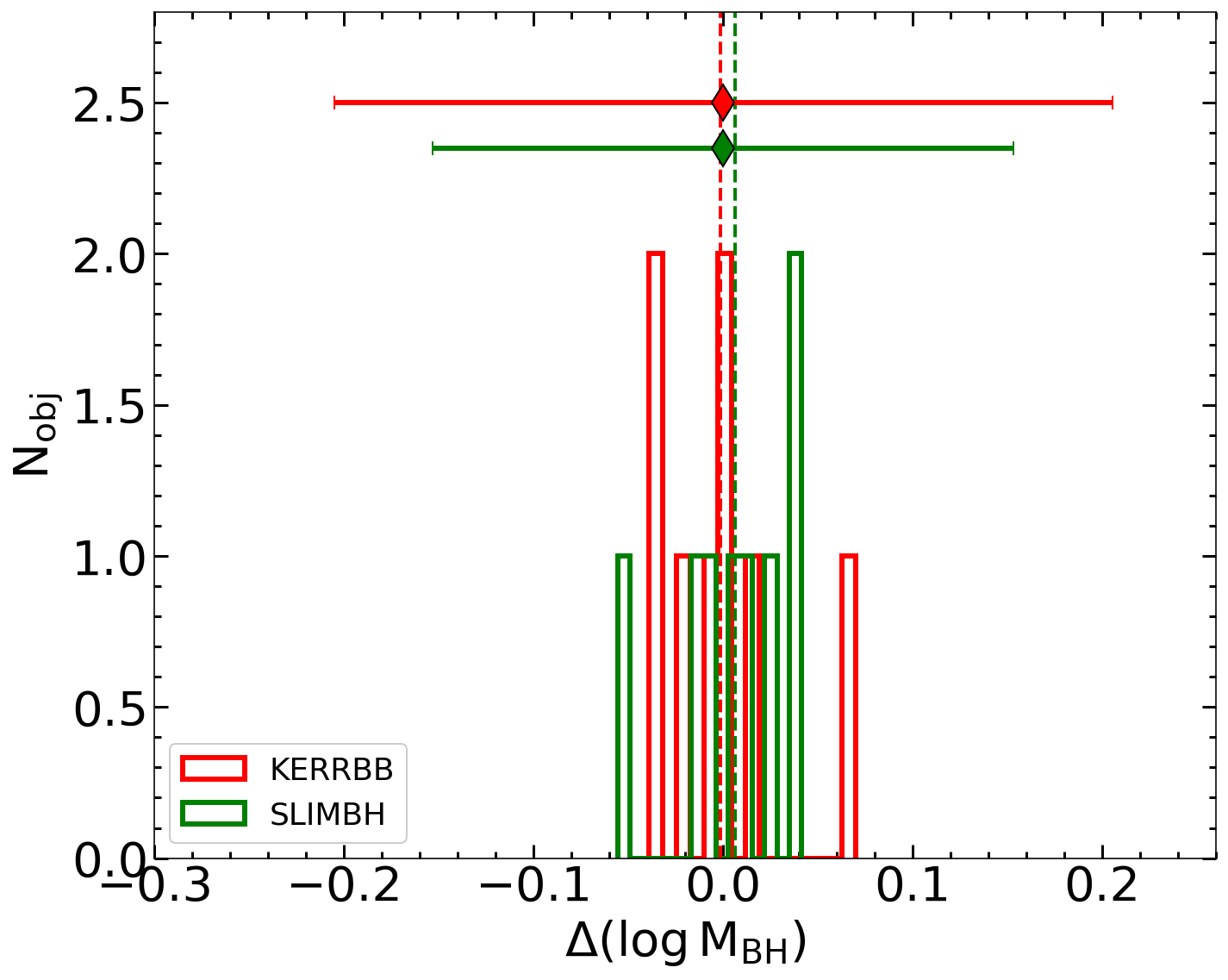}
\end{subfigure}
\hfill
\begin{subfigure}[t]{0.48\linewidth}
    \centering
    \includegraphics[width=\linewidth,clip]{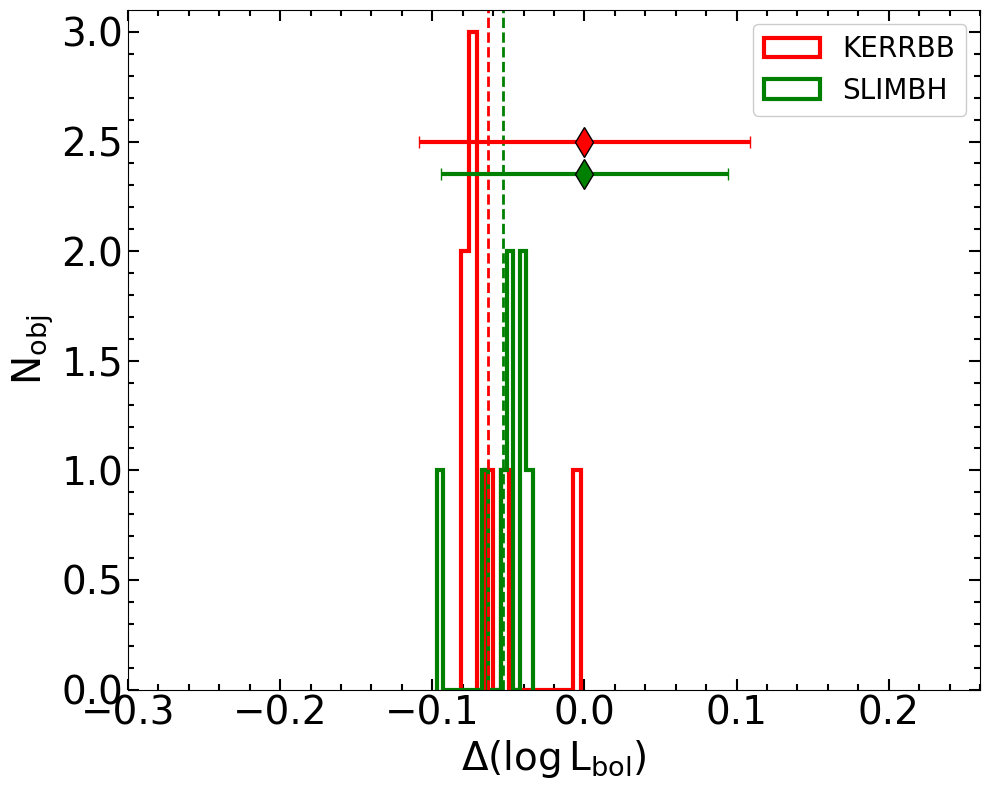}
\end{subfigure}
\caption{Distribution of the differences $\Delta \log(M_{\rm BH})$ (left panel) and $\Delta \log(L_{\rm bol})$ (right panel) between the values obtained, respectively, through our custom fitting procedure and the `BADFit' routine. 
Diamonds with error bars represent the average uncertainties on both quantities as estimated in our AD fits, while the `BADFit' ones are similar. The vertical dashed lines represent the means of the two distributions.}
\label{fig:delta_logML_comparison}
\end{figure*}

As discussed in Section \ref{sec:ad_modelling}, we also took advantage of the Python routine `BADFit' (\citealt{lai2023characterising, samuel_lai_2023_7772748}) to derive an independent estimate of both \Mbh and \Lad. This routine employs the KERRBB and SLIMBH models and performs the fit in a Bayesian framework. In Fig. \ref{fig:delta_logML_comparison} we show the distribution of the differences between the \Mbh values computed using our custom fits taking advantage of the KERRBB and SLIMBH models and those implemented in the `BADfit' routine. In particular, we define $\Delta \log(M_{\rm BH})$\,=\,$\log(M_{\rm BH, ours})$\,$-$\,$\log(M_{\rm BH, BADFit})$ and $\Delta \log(L_{\rm bol})$\,=\,$\log(L_{\rm bol, ours})$\,$-$\,$\log(L_{\rm bol, BADFit})$. The average difference for \Mbh using the KERRBB (SLIMBH) model is $-$0.001 (0.006) with a standard deviation of 0.032 (0.024). The average uncertainty on \Mbh from our KERRBB (SLIMBH) fitting routine is 0.205 (0.153). The average difference for \Lad using the KERRBB (SLIMBH) model is $-$0.063 ($-$0.053) with a standard deviation of 0.024 (0.019). The average uncertainty on \Mbh from our KERRBB (SLIMBH) fitting routine is 0.109 (0.094). Although we find a minor shift in the best-fit \Lad value from the `BADfit' routine, this is negligible with respect to the typical fit uncertainty.

\section{Composite spectra comparison}
\label{app:composite_spectra}

\begin{figure}[H]
\centering
\includegraphics[width=\linewidth,clip]{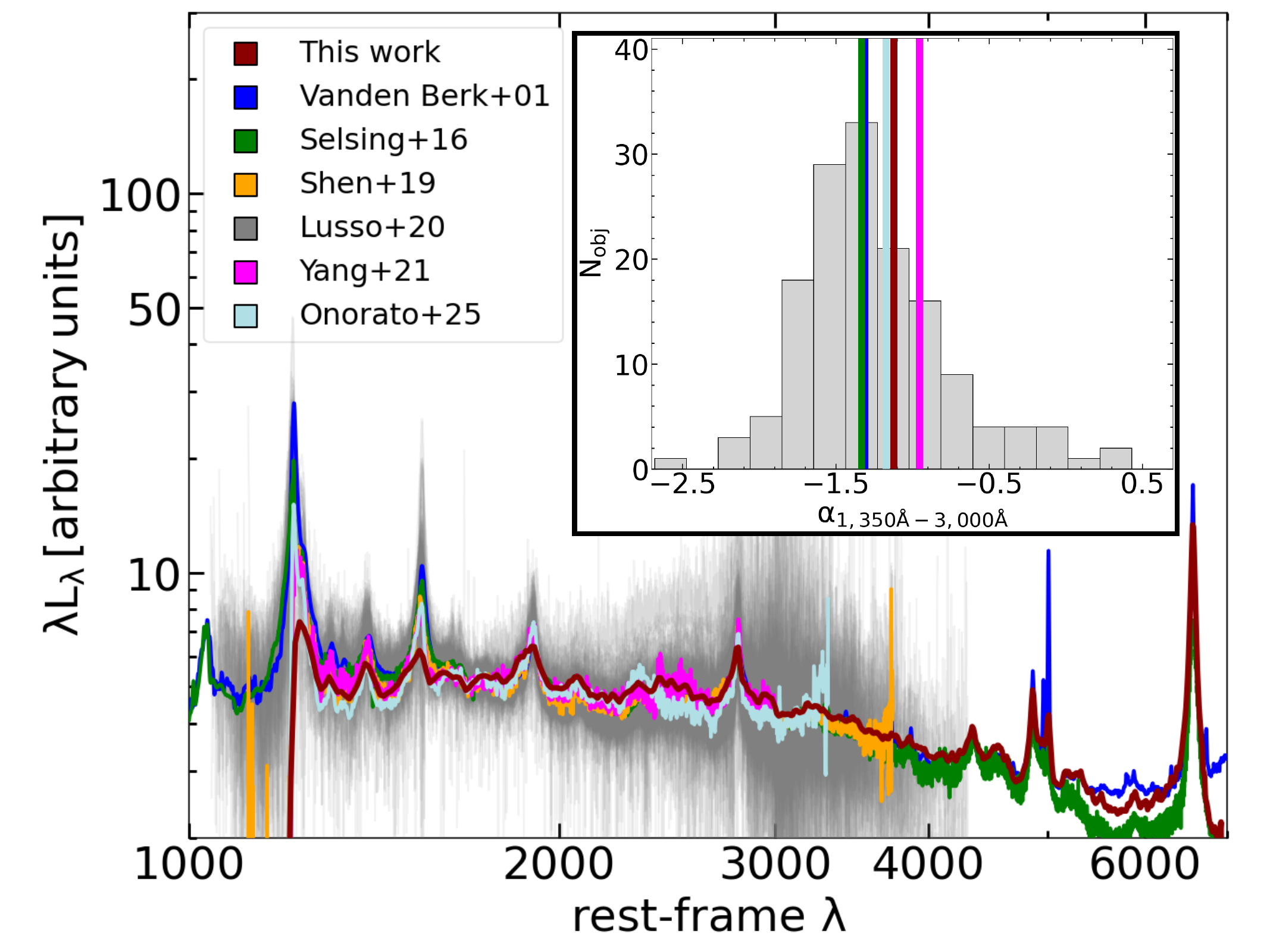}
\caption{\textit{Main panel:} Comparison between our composite spectrum and those obtained from other samples (see text). We also overplot the individual spectra of the blue QSOs from the \citet{lusso2020} sample scaled by their 1,700-\AA\ monochromatic luminosity. \textit{Inset panel:} The distribution of the 1,350--3,000~\AA\ slopes of the luminosity-matched \citet{lusso2020} sample and the other composite spectra. The slope of the \citet{shen2019gemini} and \citet{onorato2025optical} composite spectra are overlapping.}
\label{fig:stack_comparison}
\end{figure}

Here we compare the composite spectrum of the QSOs in our sample with those of other reference samples of low- and high-redshift QSOs. The procedure to build the composite spectrum of our sources follows closely that described in \citet{trefoloni2023most, trefoloni2024missing, trefoloni2024quasars}, therefore we refer the interested reader to those works for further details. The only minor difference is that, as the resolution of the fixed-slit spectra is higher than the PRISM one ($\mathcal{R}$\,$\sim$\,2,700 against $\mathcal{R}$\,$\sim$\,100), we rebinned the former to match the PRISM resolution of $\mathcal{R}$\,=\,100 at 3 $\mu$m, while preserving the total flux. In Fig. \ref{fig:stack_comparison}, we show the composite spectrum of our QSOs with those of other samples of optically selected blue QSOs (\citealt{vandenberk2001, selsing2016x, shen2019gemini, yang2021probing, onorato2025optical}). In addition, we also show individually the spectra of the $\sim$\,150 optically blue and X-ray unobscured QSOs, from the sample described in \citet{lusso2020}, falling within 0.3 dex from the mean $\log(L_{3000\,\AA}/(\rm erg \, s^{-1}))$ of our sample. We highlight that these latter sources are selected in order to minimise the possible presence of reddening via a careful UV/optical and X-ray selection (see their Section 5 for further details). The overall shape of our composite spectrum closely matches that of the other reference ones, albeit with a lower spectral resolution. This visual similarity is also quantitatively confirmed by the slope of the continuum power law joining the 1,350~\AA\, and 3,000~\AA\, monochromatic luminosities, which is close to the mean of the \citet{lusso2020} distribution, as also are those of the other composite spectra. This similarity argues in favour of our sources being fairly representative of the typical accretion state and/or the degree of reddening of bright QSOs.

\section{Emission line properties}
In this section we report the emission-line parameters measured from the spectral fits for the sample in Table \ref{tbl:spec_pars}.

\begin{table*}[h!]
\caption{Emission-line properties.}

\label{tbl:spec_pars} 
\centering
\setlength{\tabcolsep}{3pt} 
\begin{tabular}{ccccccccc} 
\hline
ID & $\log(L_{3000\,\AA})$ & $\log(L_{5100\,\AA})$ & $\sigma$ H$\alpha$ & FWHM H$\alpha$ & $\sigma$ H$\beta$ & FWHM H$\beta$ & FWHM Mg\,\textsc{ii} lor & FWHM Mg\,\textsc{ii} gau \\ 
 
& $[\rm erg \, s^{-1}]$ & $[\rm erg \, s^{-1}]$ &  [$\rm km \, s^{-1}$]	& [$\rm km \, s^{-1}$]	  & [$\rm km \, s^{-1}$]	& [$\rm km \, s^{-1}$] & [$\rm km \, s^{-1}$] & [$\rm km \, s^{-1}$] \\
\\
\hline
J0020$-$3653 & $46.36\pm0.02$ & $46.22\pm0.01$ & $2629\pm16$ & $3184\pm35$ & $3290\pm36$ & $3979\pm114$ & $2326\pm49$ & $5010\pm69$ \\
J0313$-$1806 & $46.45\pm0.05$ & $46.25\pm0.04$ & \ldots & \ldots & $2396\pm97$ & $1879\pm77$ & $3337\pm88$ & $4836\pm85$ \\
J0411$-$0907 & $46.51\pm0.01$ & $46.32\pm0.01$ & $2615\pm6$ & $2968\pm14$ & $3129\pm18$ & $3578\pm36$ & $2524\pm10$ & $6808\pm24$ \\
J0910$-$0414 & $46.64\pm0.01$ & $46.39\pm0.00$ & $3168\pm47$ & $3599\pm121$ & $3931\pm90$ & $6715\pm789$ & \ldots & \ldots \\
J1007+2115 & $46.51\pm0.05$ & $46.31\pm0.04$ & \ldots & \ldots & $3307\pm118$ & $4296\pm581$ & $3153\pm84$ & $3826\pm33$ \\
J1342+0938 & $46.29\pm0.03$ & $46.08\pm0.02$ & \ldots & \ldots & $3029\pm117$ & $3326\pm218$ & $2990\pm37$ & $4292\pm27$ \\
J1425+3254 & $46.18\pm0.01$ & $45.94\pm0.01$ & $2926\pm33$ & $2970\pm89$ & $3274\pm75$ & $4222\pm182$ & $2754\pm16$ & $5604\pm12$ \\
J2239+0207 & $45.66\pm0.01$ & $45.42\pm0.00$ & $2769\pm39$ & $2746\pm41$ & $2675\pm199$ & $2861\pm248$ & $2463\pm65^{\dagger}$ & $8538\pm296$ \\
\hline	
\end{tabular}
\tablefoot{$^{\dagger}$ As the FWHM of the Lorentzian profile falls below the nominal spectral resolution, we report the FWHM directly measured from the spectrum.}
\end{table*}

\section{SE \Mbh calibrations}

In Table \ref{tbl:MBH_calibrations}, we gather all the SE calibrations employed in this work.

\begin{table*}[h!]
\caption{SE calibrations.}
\centering
\setlength{\tabcolsep}{5pt}
\begin{tabular}{c c c c c c c c}
 \hline \noalign{\smallskip}
 Line & a & b & c & L & W & Ref. & mean (median) $\Delta \pm \sigma$\\
 (1) & (2) & (3) & (4) & (5) & (6) & (7) & (8)\\

 \hline \noalign{\smallskip}
 Mg\,\textsc{ii}       &  6.51   & 0.62 & 2.00 & $L_{3000\,\AA}$ & FWHM & \citealp[MD04]{mclure2004cosmological} & $-$0.5 ($-$0.4) $\pm$ 0.2  \\ 
 Mg\,\textsc{ii}       &  6.86   & 0.50 & 2.00 & $L_{3000\,\AA}$ & FWHM & \citealp[VO09]{vestergaard2009mass} & $-$0.4 ($-$0.3) $\pm$ 0.2 \\ 
 Mg\,\textsc{ii}       &  6.74   & 0.62 & 2.00 & $L_{3000\,\AA}$ & FWHM & \citealp[S11]{shen2011} & $-$0.3 ($-$0.2) $\pm$ 0.2 \\ 
 Mg\,\textsc{ii}       &  6.35   & 0.60 & 3.00 & $L_{3000\,\AA}$ & $\sigma$ & \citealp[S24]{shen2024sloan} & 0.2 (0.1) $\pm$ 0.3 \\ 

 \hline
 H$\beta$    &  6.91   & 0.50 & 2.00 & $L_{5100\,\AA}$ & FWHM & \citealp[VP06]{vestergaard2006determining} & $-$0.2 ($-$0.2) $\pm$ 0.3 \\ 
 H$\beta$    &  6.72   & 0.65 & 2.00 & $L_{5100\,\AA}$ & FWHM & \citealp[TN12]{trakhtenbrot2012black} & $-$0.1 ($-$0.1) $\pm$ 0.4 \\ 
 H$\beta$    &  6.94   & 0.53 & 2.00 & $L_{5100\,\AA}$ & FWHM & \citealp[B13]{bentz2013low} & $-$0.2 ($-$0.1) $\pm$ 0.3 \\ 
 H$\beta$    &  8.57   & 0.78 & 1.39 & $L_{\rm H\beta}$ & FWHM & \citealp[DB20]{db2020} & 0.2 (0.3) $\pm$ 0.3 \\ 
 H$\beta$    &  8.53   & 0.70 & 2.18 & $L_{\rm H\beta}$ & $\sigma$ & \citealp[DB20]{db2020} & 0.5 (0.5) $\pm$ 0.3 \\ 
 H$\beta$    &  6.85   & 0.50 & 2.00 & $L_{5100\,\AA}$ & $\sigma$ & \citealp[S24]{shen2024sloan} & $-$0.3 ($-$0.3) $\pm$ 0.3 \\ 

 \hline
 H$\alpha$   &  7.40   & 0.55 & 2.06 & $L_{\rm H\alpha}$ & FWHM & \citealp[GH05]{greene2005estimating} & $-$0.5 ($-$0.5) $\pm$ 0.1 \\ 
 H$\alpha$   &  8.13   & 0.46 & 2.06 & $L_{\rm H\alpha}$ & $\sigma$ & \citealp[W15]{woo2015black} & 0.1 (0.1) $\pm$ 0.2 \\ 
 H$\alpha$   &  7.73   & 0.61 & 2.00 & $L_{\rm H\alpha}$ & FWHM & \citealp[C23]{cho2023seoul} & $-$0.1 ($-$0.2) $\pm$ 0.1 \\ 
 H$\alpha$   &  8.18   & 0.81 & 1.63 & $L_{\rm H\alpha}$ & FWHM & \citealp[DB25]{dalla2025estimating} & 0.3 (0.3) $\pm$ 0.2 \\ 
 H$\alpha$   &  7.90   & 0.55 & 2.61 & $L_{\rm H\alpha}$ & $\sigma$ & \citealp[DB25]{dalla2025estimating} & 0.2 (0.2) $\pm$ 0.2 \\ 

 \hline
\end{tabular}
\tablefoot{Columns represent respectively: line employed (1), coefficients of the calibration as expressed in Eq.~\ref{eq:Mbh_vir_eq} (2,3,4), continuum or line luminosity used by the calibration (5), proxy of the line width (6), reference (7), and mean (median) deviation between the AD and SE masses $\pm$ standard deviation.}
\label{tbl:MBH_calibrations}
\end{table*}

\section{Fit atlas}
\label{app:fit_atlas}
In Fig.~\ref{fig:fit_atlas} we present an atlas of all the spectral and AD fits to the QSOs in our sample.

\begin{figure*}[h!]
\centering
\includegraphics[width=1.0\linewidth,clip]{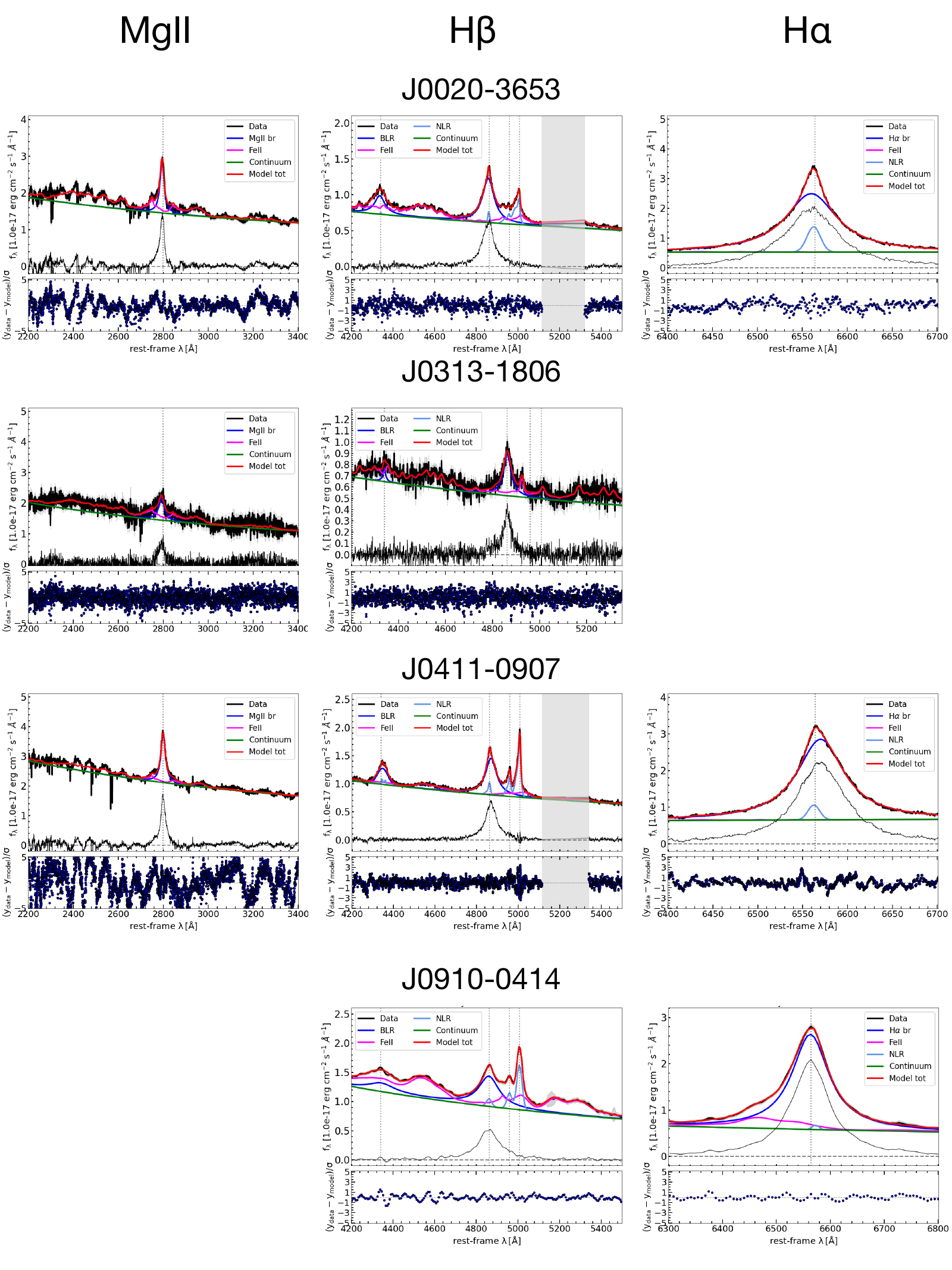}
\caption{Spectral and AD fits of the QSOs in our sample.}
\label{fig:fit_atlas}
\end{figure*}
\newpage

\addtocounter{figure}{-1}
\begin{figure*}[h!]
\centering
\includegraphics[width=1.0\linewidth,clip]{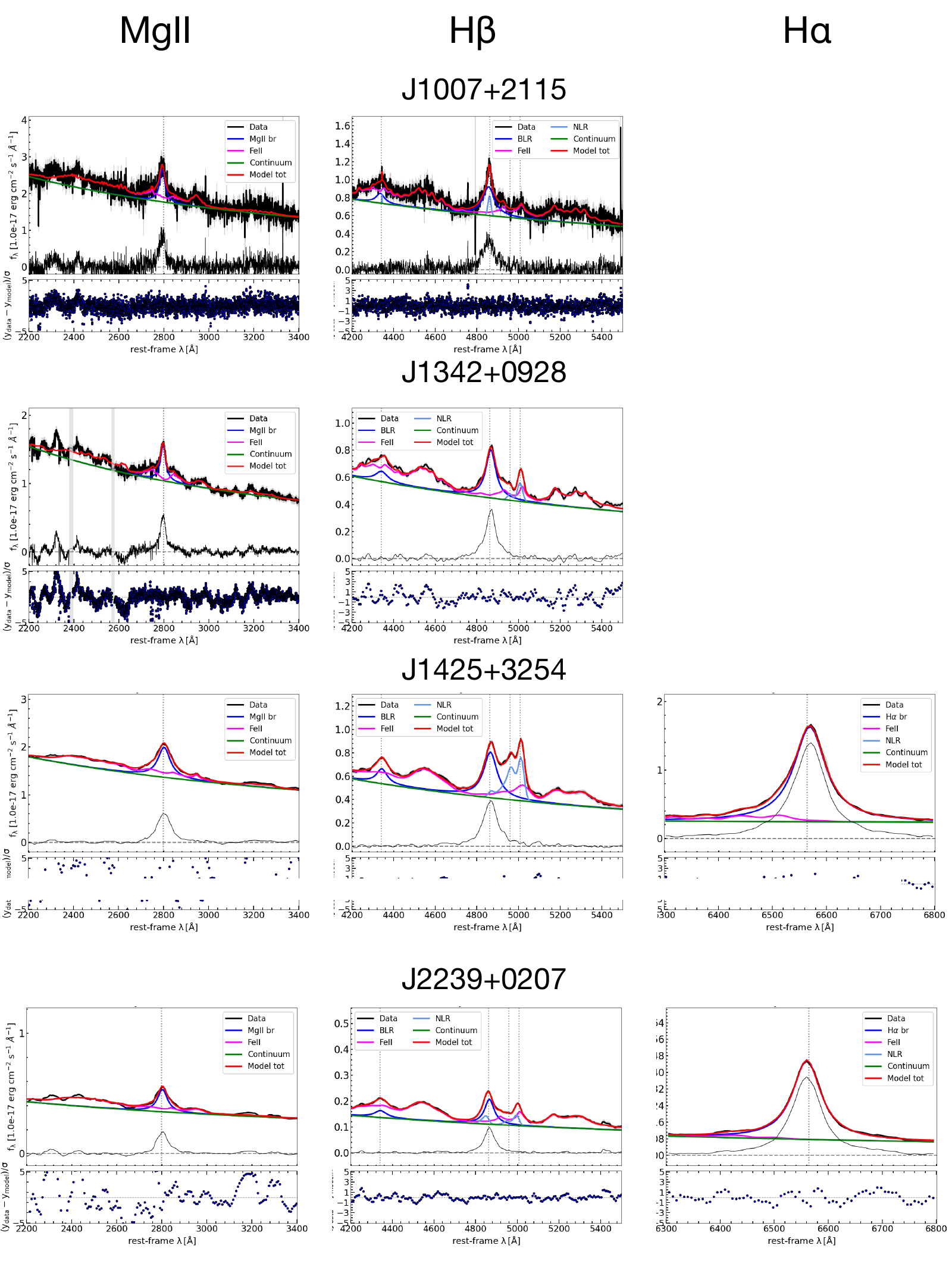}
\caption{continued.}
\end{figure*}
\newpage

\addtocounter{figure}{-1}
\begin{figure*}[h!]
\centering
\includegraphics[width=1.0\linewidth,clip]{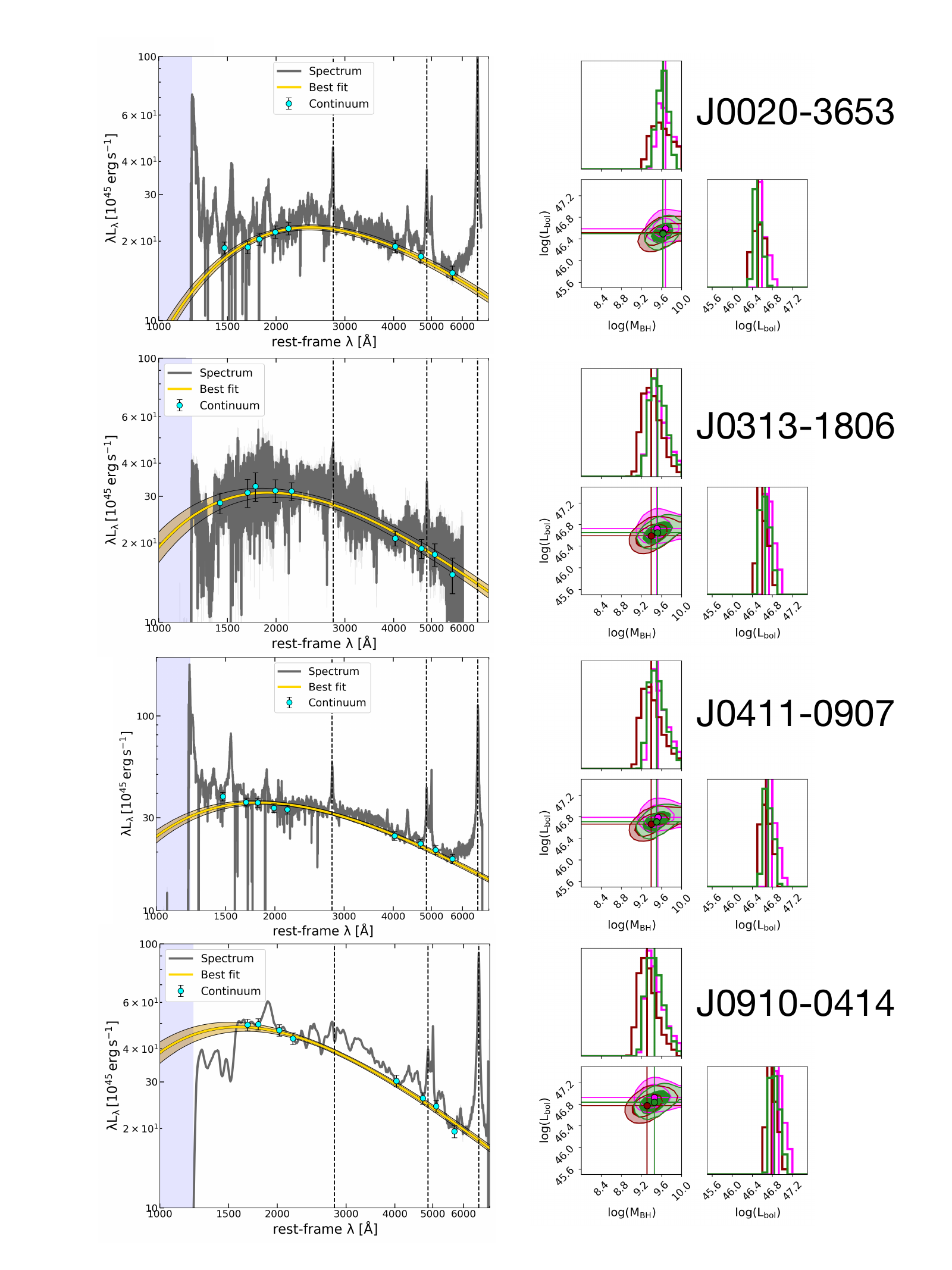}
\caption{continued.}
\end{figure*}
\newpage

\addtocounter{figure}{-1}
\begin{figure*}[h!]
\centering
\includegraphics[width=1.0\linewidth,clip]{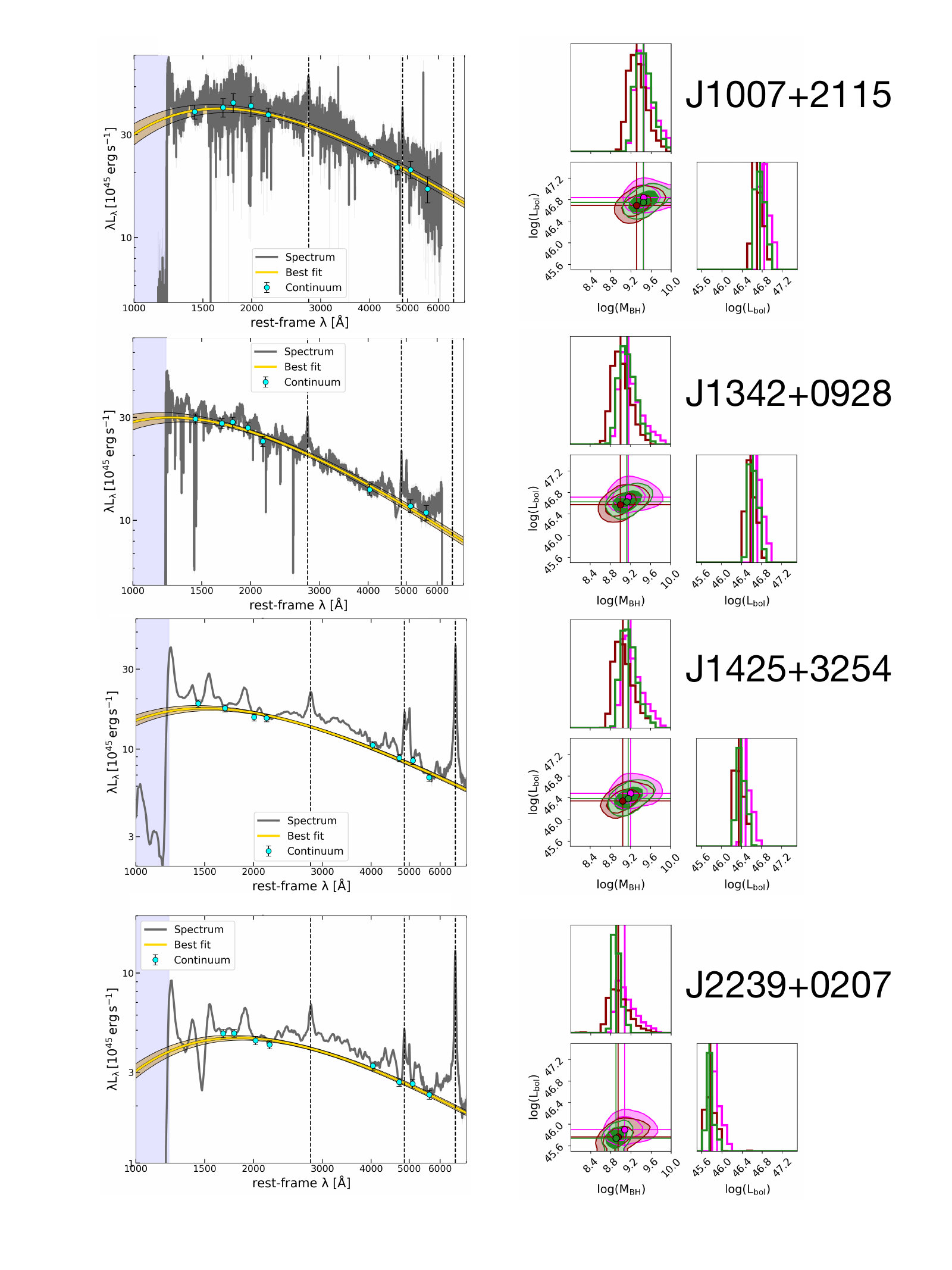}
\caption{continued.}
\end{figure*}
\newpage

\end{appendix}

\end{document}